\pdfoutput=1
\documentclass[10pt]{jpaper}

\usepackage{etex}
\usepackage[sort,nocompress,adjust]{cite} % autosort groups of citations
\usepackage{amsmath,amssymb,amsfonts}
\usepackage{algorithmic}
\usepackage{graphicx}
\usepackage{listings}
\usepackage{xcolor}
\definecolor{olive}{rgb}{0.33, 0.42, 0.18}
\usepackage{makecell}
\usepackage{textcomp}
\usepackage{fancyhdr}
\usepackage{tabu}
\usepackage{enumitem}
\usepackage[normalem]{ulem}
\usepackage{inconsolata}
\usepackage[scaled]{beramono}
\usepackage{float}
\usepackage{caption}

\def\BibTeX{{\rm B\kern-.05em{\sc i\kern-.025em b}\kern-.08em
T\kern-.1667em\lower.7ex\hbox{E}\kern-.125emX}}

\usepackage{verbatim}   % allows multi-line comments

\usepackage{wrapfig}
\usepackage{multirow}
\usepackage{balance}
\usepackage{cite}

\usepackage{pifont}

\usepackage{rotating}

\usepackage{arydshln}

\usepackage{placeins}

\usepackage{gensymb}

\usepackage{enumitem}
\usepackage[ruled,vlined]{algorithm2e}
\usepackage{upgreek,textgreek}
\usepackage[htt]{hyphenat}
\usepackage{appendix}

\hypersetup{colorlinks,
	linkcolor=blue,
	citecolor=blue,
	urlcolor=blue,
	filecolor=blue}

\usepackage{mathtools}

\usepackage{tikz}

\usepackage{listings}
\usepackage[utf8]{inputenc}
\lstdefinestyle{customcpp}{
	 aboveskip=0in,
	  belowskip=0in,
	   abovecaptionskip=0in,
	    belowcaptionskip=0in,
	     captionpos=b,
	      xleftmargin=\parindent,
	       language=C++,
	        morekeywords={forall},
		 showstringspaces=false,
		  basicstyle={\linespread{0.6}\fontseries{sb}\small\ttfamily},
		   keywordstyle=\bfseries,
		    commentstyle=\itshape\color{green!40!black},
	    }

\begin{document}

%
% The "title" command has an optional parameter, allowing the author to define a "short title" to be used in page headers.
%\title{Sage: Practical \& Scalable ML-Driven Performance Debugging in Microservices\vspace{-0.16in}}
\title{Sinan: Data-Driven, QoS-Aware Cluster Management for Microservices\vspace{-0.16in}}
%
% The "author" command and its associated commands are used to define the authors and their affiliations.
% Of note is the shared affiliation of the first two authors, and the "authornote" and "authornotemark" commands
% used to denote shared contribution to the research.
\author{Yanqi Zhang, Weizhe Hua, Zhuangzhuang Zhou, Edward Suh, and Christina Delimitrou\\ Cornell University\\Contact author: yz2297@cornell.edu}

\date{}

\maketitle
\pagestyle{plain}

\begin{abstract}
Cloud applications are increasingly shifting from large monolithic services, to large numbers of loosely-coupled, 
specialized microservices. Despite their advantages in terms of facilitating development, deployment, modularity, and isolation, 
microservices complicate resource management, as dependencies between them introduce backpressure effects and 
cascading QoS violations. 

We present Sinan, a data-driven cluster manager for interactive cloud microservices that is online and QoS-aware. 
Sinan leverages a set of scalable and validated machine learning models to determine the performance impact of dependencies 
between microservices, and allocate appropriate resources per tier in a way that preserves the end-to-end tail latency target. 
We evaluate Sinan both on dedicated local clusters and large-scale deployments on Google Compute Engine (GCE) across representative 
end-to-end applications built with microservices, such as social networks and hotel reservation sites. We show that 
Sinan always meets QoS, while also maintaining cluster utilization high, in contrast to prior work which leads to % results 
unpredictable performance or sacrifices resource efficiency. Furthermore, the techniques in Sinan are explainable, meaning that 
cloud operators can yield insights from the ML models on how to better deploy and design their applications to reduce unpredictable performance. 

\end{abstract}

\section{Introduction}
\label{sec:intro}

In recent years, cloud applications have progressively shifted from \textit{monolithic} services to graphs with hundreds of single-purpose and loosely-coupled 
\textit{microservices}~\cite{suresh2017distributed,gan2019open,Gan18,gan2019open,Gangan2018seer,sriraman2018mu,Cockroft15,Cockroft16,twitter_decomposing}. 
This shift is becoming increasingly pervasive, with large cloud providers, such as Amazon, Twitter, Netflix, 
and eBay having already adopted this application model~\cite{Cockroft15,Cockroft16,twitter_decomposing}. 
%\begin{wrapfigure}[13]{l}{0.28\textwidth}
%	\centering
%	\includegraphics[scale=0.28, trim=0 0.2cm 0 5.4cm, clip=true]{figures/motivation_comparison3.pdf}
%	\caption{\label{fig:motivation} \figurecaptionsize{Differences in the design and deployment of monoliths and microservices.} }
%\end{wrapfigure}

Despite several advantages, such as modular and flexible development and rapid iteration, microservices also introduce 
new system challenges, especially in resource management, since the complex topologies of microservice dependencies 
exacerbate queueing effects, and introduce cascading Quality of Service (QoS) violations that are difficult to identify and correct in a timely manner~\cite{gan2019open,zhou2018overload}. 
Current cluster managers are designed for monolithic applications or applications consisting of a few pipelined tiers, and 
are not expressive enough to capture the complexity of microservices~\cite{Borg,Lin11,Meisner11,Lo14,Lo15,gan2019open,Ousterhout13,omega13,Cloudscale,Delimitrou13,Delimitrou14,Delimitrou15,Delimitrou16,Delimitrou13e,Delimitrou14b,Delimitrou17}.
Given that an increasing number of production cloud services, such as EBay, Netflix, Twitter, and Amazon, are now designed as microservices, 
addressing their resource management challenges is a pressing need~\cite{Cockroft15,Cockroft16,gan2019open}. 

We take a data-driven approach to tackle the complexity microservices introduce to resource management. Similar machine learning (ML)-driven approaches 
have been effective at solving resource management problems for large-scale systems 
in previous work~\cite{Delimitrou14,cortez2017resource,rzadca2020autopilot,Delimitrou14b,Delimitrou13,Delimitrou17,Delimitrou14b,Delimitrou13e}. Unfortunately, 
these systems are not directly applicable to microservices, as they were designed for monolithic services, and hence do not account for the impact of dependencies between 
microservices on end-to-end performance. 

We present Sinan, a scalable and QoS-aware resource manager for interactive cloud microservices. 
Instead of tasking the user or cloud operator with inferring the impact of dependencies between microservices, 
Sinan %introduces an ML-based approach to microservice management that 
leverages a set of validated ML models 
to automatically determine the impact of per-tier resource allocations on end-to-end performance, 
and assign appropriate resources to each tier. 

Sinan first uses an efficient space exploration algorithm to examine the space of possible resource allocations, 
especially focusing on corner cases that introduce QoS violations. This yields a training dataset used to train two models: 
a Convolutional Neural Network (CNN) model for detailed short-term performance prediction, and a Boosted Trees model that evaluates 
the long-term performance evolution. The combination of the two models allows Sinan to both examine 
the near-future outcome of a resource allocation, % in detail, % in a detailed way, 
and to account for the system's inertia 
in building up queues with higher accuracy than a single model examining both time windows. Sinan operates online, adjusting per-tier resources dynamically according to the 
service's runtime status and end-to-end QoS target. Finally, Sinan is implemented 
as a centralized resource manager with global visibility into the cluster and application state, and 
with per-node resource agents that track per-tier performance and resource utilization. 

We evaluate Sinan using two end-to-end applications from DeathStarBench~\cite{gan2019open}, built with interactive microservices: 
a social network and a hotel reservation site. We compare Sinan against both traditionally-employed 
empirical approaches, such as autoscaling~\cite{aws_step_scaling}, and previous research on multi-tier service scheduling based on queueing analysis, 
such as PowerChief~\cite{powerchief}. We demonstrate that Sinan outperforms previous work both in terms of performance and resource efficiency, 
successfully meeting QoS for both applications under diverse load patterns.
On the simpler hotel reservation application, Sinan saves \textit{25.9\%} on average, and up to \textit{46.0\%} of the amount of resources used by other QoS-meeting methods. 
On the more complex social network service, where abstracting application complexity is more essential, Sinan saves \textit{59.0\%} of resources on average, and up to \textit{68.1\%}, 
essentially accommodating twice the amount of requests per second, without the need for more resources. 
We also validate Sinan's scalability through large-scale experiments on approximately 100 container instances on Google Compute Engine (GCE), and demonstrate that the models deployed on the local cluster can be reused on GCE with only minor adjustments instead of retraining. 
%while being robust against skews in user workloads, requiring a small amount of data by fine tuning models trained locally.

Finally, we demonstrate the explainability benefits of Sinan's models, delving into the insights they can provide for the design of large-scale systems. 
Specifically, we use an example of Redis's log synchronization, which Sinan helped identify as the source 
of unpredictable performance out of tens of dependent microservices to show that the system %systems like Sinan %show that ML-driven approaches 
can offer practical and insightful solutions for clusters whose scale make previous empirical approaches impractical.

\section{Overview}
\label{sec:overview}
% \vspace{-0.08in}

\subsection{Problem Statement}

Sinan aims to manage resources for complex, interactive microservices with tail latency 
QoS constraints in a scalable and resource-efficient manner. 
Graphs of dependent microservices typically include tens to hundreds of tiers, each with 
different resource requirements, scaled out and replicated for performance and reliability. 
%with different resource requirements and limitations for each. 
Section~\ref{sec:applications} describes some motivating examples of such services with diverse functionality used in this work; 
other similar examples can be found in~\cite{Cockroft15,Cockroft16,twitter_decomposing,sriraman2018mu}. 

%The complexity and scale of microservices-based applications make resource management challenging, especially given the fact that dependent tiers are not perfect pipelines, and hence can introduce backpressure effects that are hard to detect and prevent~\cite{Delimitrou19,Gan19,gan2018seer}. 
Most cluster managers focus on CPU and memory management~\cite{rzadca2020autopilot,cortez2017resource,Borg}. 
Microservices are by design mostly stateless, hence their performance is defined by their CPU allocation. Given this, Sinan primarily focuses 
on allocating CPU resources to each tier~\cite{gan2019open}, both at sub-core and multi-core granularity, leveraging Linux cgroups through the Docker API~\cite{docker}. 
%In this paper, the cpu allocation is enforced by cpu limit api of docker, which is implemented with linux cgroup. 
We also provision each tier with the maximum profiled memory usage to eliminate out of memory errors.

% \vspace{-0.08in}
\subsection{Motivating Applications}
\label{sec:applications}

We use two end-to-end interactive applications from DeathStarBench~\cite{gan2019open}: 
a hotel reservation service, and a social network. 
%such as {\texttt{NGINX}}~\cite{nginx}, {\texttt{memcached}}~\cite{memcached}, {\texttt{MongoDB}}~\cite{mongodb} and {\texttt{RabbitMQ}}~\cite{rabbitmq}, and includes 27 and 17 tiers correspondingly. The rest of the section overviews the design and functionality of the three applications. 

\subsubsection{Hotel Reservation}
\begin{figure}[h!]
 \centering
   \vspace{-0.10in}
   \includegraphics[scale=0.33,trim=0.3 1cm 2cm 3.8cm,clip=true]{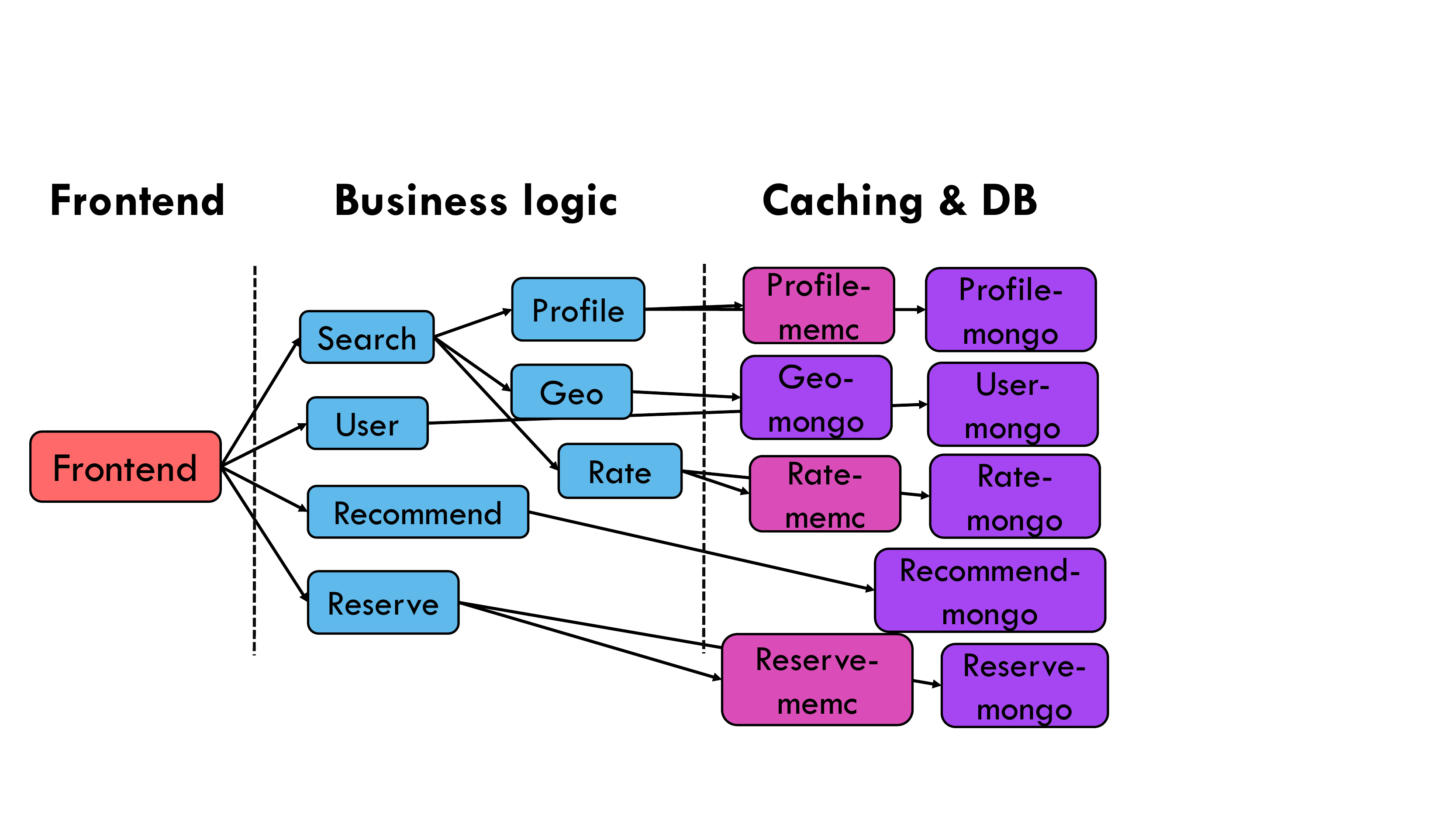}
   \vspace{-0.30in}
	 \caption{Hotel reservation microservice architecture~\cite{gan2019open}. \textcolor{black}{Client requests first reach a front-end webserver,
        and, depending on the type of requests, are then directed to logic tiers implementing functionality for searching hotels, completing hotel reservations, and getting recommendations on available hotels. At the right-most of the figure,
	 the requests reach the back-end databases, implemented both with in-memory caching tiers (memcached), and persistent databases (MongoDB). }}
   \label{fig:hotel_reserv}
   \vspace{-0.08in}
 \end{figure}

The service is an online hotel reservation site, whose architecture is shown in Fig.~\ref{fig:hotel_reserv}. 

\noindent{\bf{Functionality: }}The service supports searching for hotels using geolocation, placing reservations, and getting recommendations. 
It is implemented in Go, and tiers communicate over gRPC~\cite{grpc}. 
Data backends are implemented in {\texttt{memcached}} for in-memory caching, 
and {\texttt{MongoDB}}, for persistent storage. The database is populated with 80 hotels and 500 active users. 

% A request is directed 
% to a front-end webserver, and depending on the request type, it is then forwarded to downstream services including search, 
% reservation and recommendation engines, which allow users to explore hotel availability in a given region and place a reservation, 
% or get recommendations based on metrics, such as geolocation, price and availability.

% \vspace{0.05in}
\subsubsection{Social Network}
\begin{figure}[h!]
 \centering
% %   \includegraphics[scale=0.15,viewport=260 30 700 400]{figures/social_network.pdf}
   \includegraphics[scale=0.30,trim=0.4cm 1.8cm 1cm 0.4cm,clip=true]{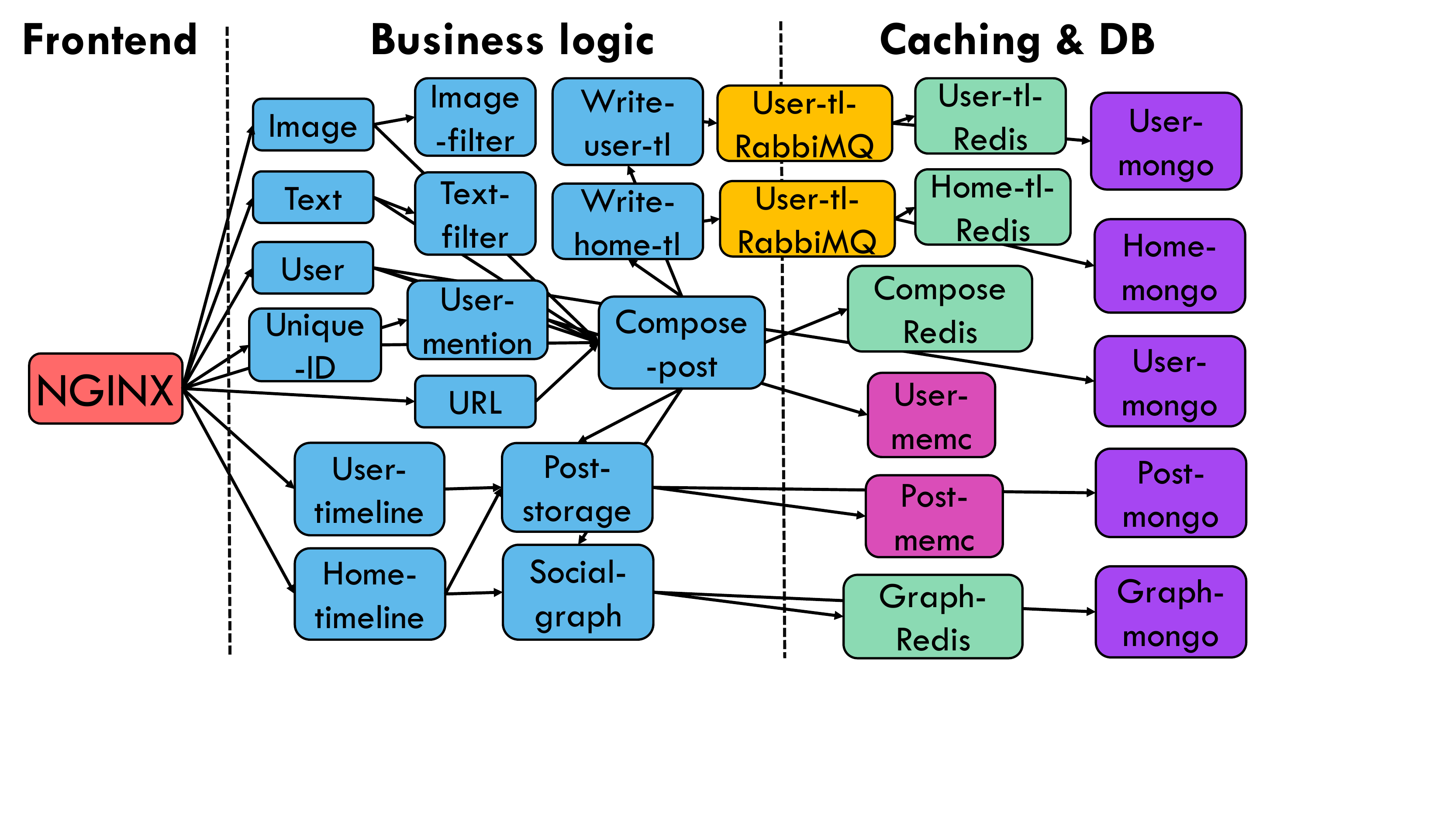}
   \vspace{-0.36in}
	 \caption{Social Network microservice architecture~\cite{gan2019open}. \textcolor{black}{Client requests first reach Nginx, which works as frontend http servers. Then, depending on the type of user request, a number of logic, mid-tiers will
        be invoked to create a post, read a user's timeline and to follow/unfollow users. At the right-most of the figure,
	 the requests reach the back-end databases, implemented both with in-memory caching tiers (memcached and Redis), and persistent databases (MongoDB). }}
   \label{fig:social_network}
   \vspace{-0.08in}
 \end{figure}

The end-to-end service implements a broadcast-style social network with uni-directional follow relationships, shown in Fig.~\ref{fig:social_network}. 
Inter-microservice messages use Apache Thrift RPCs~\cite{thrift}. 

\noindent{\bf{Functionality: }} Users can create posts embedded with text, media, links, and tags to other users, which are then broadcasted to all their followers. 
The texts and images uploaded by users, specifically, go through image-filter (a CNN classifier) and text-filter services (an SVM classifier), 
and contents violating the service's ethics guidelines are rejected. Users can also read posts on their timelines. 
We use the {\texttt{Reed98}}~\cite{nr} social friendship network to populate the user database. 
User activity follows the behavior of Twitter users 
reported in~\cite{kwak2010twitter}, and the distribution of post text length emulates Twitter's text length distribution~\cite{gligoric2018constraints}. 

%Block size represents the initial over-provisioned core allocation per tier.

% Inter-microservice messages use Apache Thrift RPCs~\cite{thrift}.
% The service's backend uses {\texttt{memcached}}~\cite{memcached} for caching, and {\texttt{MongoDB}}~\cite{mongodb} for persistently storing posts, user profiles,
% and media. Index information, such as user timeline indices, are stored in {\texttt{Redis}}. {\texttt{RabbitMQ}}~\cite{rabbitmq} instances 
% are added between business logic tiers and {\texttt{MongoDB}} to make time-consuming database write operations asynchronous, and prevent them from blocking upstream connections. 
% {\texttt{Reed98}} is a friendship graph extracted from Facebook, and consists of 962 people and 18.1K edges, 
% where an edge is a follow relationship. 
% , where a user's posting activity positively correlates with the number 
% of their followers.
%In our \texttt{Social Network} deployment, we assume that the social friendship graph is rarely changed and focus on measuring performance of reading and composing posts. 
% \vspace{-0.08in}

%- what problem are we solving

%Move to end of challenges - why is ML suitable? 

% \vspace{-0.04in}
\subsection{Management Challenges \& the Need for ML}
\label{sec:challenges}

Resource management in microservices faces four challenges. 

%the overhead of collecting data to train the ML models used by the scheduler as well (see Sec.~\ref{sec:design_models}). Specifically, for the ML model to ensure high decision quality, the training dataset needs to cover a sufficient spectrum of application behaviors. Randomly exploring the resource allocation space during data collection would result in only exploring a tiny subset of the entire space. For the \textit{Social Network} application specifically, collecting data every 1 second for 24 hours only covers $4.45\times10^{-34}$ of the entire space, even without considering that identical resource actions can result in different performance under different system states. Sec.~\ref{sec:data_collection} describes how Sinan handles data collection to ensure high coverage of interesting resource behaviors. 

\vspace{0.08in}
\noindent{\bf{1. Dependencies among tiers}} 
Resource management in microservices is additionally complicated by the fact that dependent microservices are not perfect pipelines, 
and hence can introduce backpressure effects that are hard to detect and prevent~\cite{gan2019open,zhou2018overload}. %Dependencies often exist among individual microservices, as multiple tiers come together to implement the end-to-end service. 
These dependencies can be further exacerbated by the specific RPC and data store API implementation. 
%Dependencies complicate identifying bottleneck tiers, as shown by the backpressure effect described in~\cite{gan2019open}. 
Therefore, the resource scheduler should have a global view of the microservice graph and be able to anticipate the impact of dependencies on end-to-end performance.

\vspace{0.08in}
\noindent{\bf{2. System complexity}} 
Given that application behaviors change frequently, resource management decisions
need to happen online. This means that the resource manager must traverse a space 
that includes all possible resource allocations per microservice in a practical manner. 
%In resource management and any stochastic decision making problem in general, decisions are made based on the system state. 
Prior empirical approaches use resource utilization~\cite{aws_step_scaling}, or latency measurements~\cite{chen2019parties, Delimitrou14, Lo14} 
to drive allocation decisions. %as the input driving decision making. 
Queueing approaches similarly characterize the system state using queue lengths~\cite{powerchief}. 
Unfortunately these approaches cannot be directly employed in complex microservices with tens of dependent tiers. % and complex dependencies. 
%diverse resource usage bebahvior, and the system state can hardly be represented with utilization of a single resource. Moreover, 
First, microservice dependencies mean that resource usage across tiers is codependent, so examining fluctuations in individual tiers can attribute poor performance to 
the wrong tier. %point to the wrong source of poor performance. %\note{I don't understand this point. You use resource and latency metrics in Sinan later on; they are from multiple tiers but they are similar metrics to what prior techniques use. I think you're opening yourself up to criticism here. }
Similarly, although queue lengths are accurate indicators of a microservice's system state, obtaining exact queue lengths is hard. % for two reasons. 
First, queues exist across the system stack from the NIC and OS, to the network stack and application. Accurately tracking queue lengths requires application changes and 
heavy instrumentation, which can negatively impact performance and/or is not possible in public clouds. Second, the application may 
include third-party software whose source code cannot be instrumented. 
%In the end, resource managers need to formalize system state that incorporates the dependencies and different behaviors of tiers, and is precise enough to make correct resource allocation decisions, with lightweight albeit indirect information, like per-tier resource utilization.
Alternatively, expecting the user to express each tier's resource sensitivity is problematic, as users
already face difficulties correctly reserving resources for simple, monolithic workloads, leading to well-documented underutilization~\cite{Delimitrou14,GoogleTrace},
%provide the information on each tier's resource sensitivity is also problematic,
and the impact of microservice dependencies is especially hard to assess, even for expert developers.

\begin{figure}
%\vspace{-0.18in}
    \centering
    \begin{tabular}{@{}c@{}}
     \vspace{-0.14in}
    \includegraphics[scale=0.26,trim=0.2cm 0 25cm 0cm,clip=true]{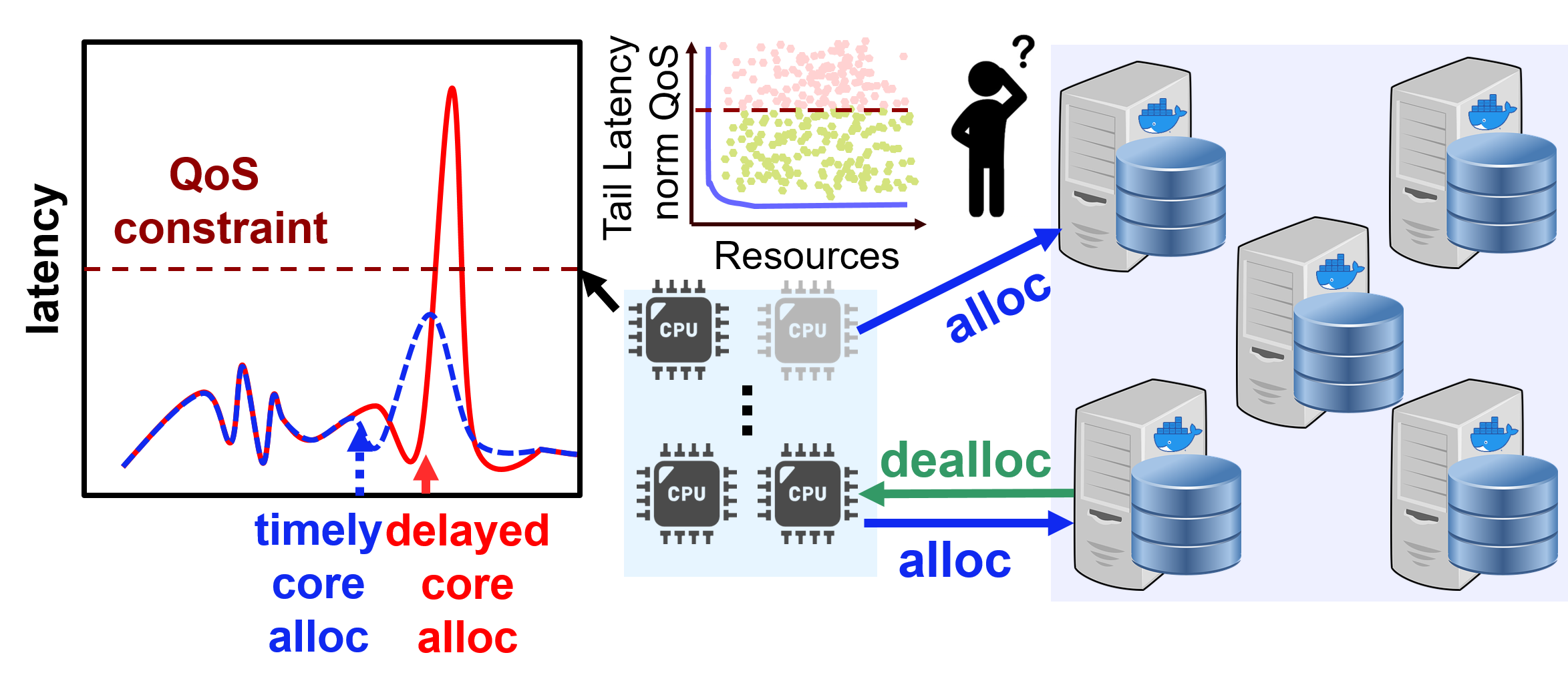}
    \end{tabular}\\
     %\vspace{-0.14in}
	\caption{\label{fig:challenges}The figure showcases the delayed queueing effect in microservices; QoS violations 
  that are not detected eagerly (blue line), become unavoidable (red), even if later action is taken. }
\end{figure}

\vspace{0.08in}
\noindent\textbf{3. Delayed queueing effect} Consider a queueing system with processing throughput $T_o$ under a latency QoS target, like the one in Fig.~\ref{fig:challenges}. 
$T_o$ is a non-decreasing function of the amount of allocated resources $R$. For input load $T_i$, $T_o$ should equal or slightly 
surpass $T_i$ for the system to stably meet QoS, while using the minimum amount of resources $R$ needed. 
Even when $R$ is reduced, such that $T_o < T_i$, QoS will not be immediately violated, since queue accumulation takes time. 
% \begin{wrapfigure}[14]{r}{0.16\textwidth}
% \centering
%   \vspace{-0.4in}
% 	\hspace*{-0.08in}\includegraphics[scale=0.21,trim=0.2cm 0 25cm 0cm,clip=true]{figures/challenge.png}
%   \vspace{-0.35in}
%   \caption{Delayed queueing effect; QoS violations 
%   not detected eagerly (blue line), become unavoidable (red). }
% %   \vspace{0.15in}
%   \label{fig:challenges}
% \end{wrapfigure}

%\begin{wrapfigure}[20]{r}{0.26\textwidth}%[!h]

The converse is also true; by the time QoS is violated, the built-up queue takes a long time to drain, even if resources are upscaled 
immediately upon detecting the violation (red line). Multi-tier microservices are complex queueing systems with queues both across 
and within microservices. This delayed queueing effect highlights the need for ML to evaluate the long-term impact of resource allocations, %management actions, 
and to proactively prevent the resource manager from reducing resources too aggressively, to avoid latency spikes with long recovery periods. 
To mitigate a QoS violation, % caused by deallocating resources, 
the manager must increase resources proactively (blue line), otherwise the violation becomes unavoidable, even if more resources are allocated a posteriori.

%The complexity and scale of microservices-based applications make resource management challenging, especially given the fact that dependent tiers are not perfect pipelines, and hence can introduce backpressure effects that are hard to detect and prevent~\cite{Delimitrou19,Gan19,gan2018seer}.

\vspace{0.08in}
\noindent{\bf{4. Boundaries of resource allocation space}} 
Data collection or profiling are essential to the performance of any model. 
Given the large resource allocation space in microservices, it is essential for any resource manager 
to quickly identify the boundaries of that space that allow the service to meet its QoS, with the minimum resource amount~\cite{denning1968working}, 
so that neither performance nor resource efficiency are sacrificed. 
Prior work often uses random exploration of the resource space~\cite{Delimitrou14,Lo14,chen2019parties} or uses prior system state as the training dataset~\cite{gan2018seer}. 
Unfortunately, while these approaches work for simpler applications, in microservices they are prone to covariant shift. 
Random collection blindly explores the entire space, even though many of the explored points may never occur during the system's normal operation, 
and may not contain any points close to the resource boundary of the service. On the contrary, data from operation logs are biased 
towards regions that occur frequently in practice but similarly may not include points close to the boundary, as cloud systems often overprovision resources to 
ensure that QoS is met. 
%which indicate operation that \textit{just} meets QoS with the minimum resource allocation. In order to improve the performance of resource managers and 
To reduce exploration overheads it is essential for a cluster manager to efficiently examine the necessary and sufficient number of points in the resource 
space that allow it to \textit{just} meet QoS with the minimum resources. 
%, efficient space exploration algorithm that is carefully guided to explore the region on the edge of working set is needed for data collection.

% \vspace{-0.08in}
\subsection{Proposed Approach}

These challenges suggest that empirical resource management, such as autoscaling~\cite{aws_step_scaling} or queueing analysis-based approaches 
for multi-stage applications, such as PowerChief~\cite{powerchief}, are prone to unpredictable performance and/or resource inefficiencies. 
To tackle these challenges, we take a data-driven approach that abstracts away the complexity of microservices from the user, 
and leverages ML to identify the impact of dependencies on end-to-end performance, and make allocation decisions. %Specifically, our contribution consists of two parts: 
%+and leverages ML to identify the impact of dependencies on end-to-end performance, and make allocation decisions.
We also design an efficient space exploration algorithm that explores the resource allocation space, especially boundary regions that may introduce QoS violations, for different application scenarios.
Specifically, Sinan's ML models predict the end-to-end latency and the probability of a QoS violation for a resource configuration,
given the system's state and history. The system uses these predictions to maximize resource efficiency, while meeting QoS.
%\begin{itemize}
%     \item \textbf{Efficient boundary-aware space exploration} We design an efficient space exploration algorithm that explores the resource allocation space, especially boundary regions that may introduce QoS violations, for different application scenarios.
%     \item \textbf{Hybrid ML model for proactive resource allocation}  We design a hybrid ML model that predicts the end-to-end latency and the probability of a QoS violation for a resource configuration, given the system's state and history. The system uses these predictions to maximize resource efficiency, while meeting QoS. 
%   \end{itemize}

%Below, we first %introduce motivating applications (Section~\ref{sec:applications}), and then 
%describe Sinan's ML models (Section~\ref{sec:design_models}), and then Sinan's system architecture (Section~\ref{sec:design}). 

{\color{black} At a high level, the workflow of Sinan is as follows: the data collection agent collects training data, 
using a carefully-designed algorithm which addresses Challenge 4 (efficiently exploring the resource space). With the collected data, 
Sinan trains two ML models: a convolution neural network (CNN) model and a boosted trees (BT) model. The CNN handles Challenges 1 and 2 
(dependencies between tiers and navigating the system complexity), by predicting the end-to-end tail latency in the near future. The BT model 
addresses Challenge 3 (delayed queueing effect), by evaluating the probability for a QoS violation further into the future, to account for the system's inertia in building up queues. 
At runtime, Sinan infers the instantaneous tail latency and the probability for an upcoming QoS violation, and adjusts resources accordingly to satisfy the QoS constraint. 
If the application or underlying system change at any point in time, Sinan retrains the corresponding models to account for the impact of these changes on end-to-end performance. }

\section{Machine Learning Models}
\label{sec:design_models}

\textcolor{black}{The objective of Sinan's ML models is to accurately predict the performance of the application given a certain resource allocation. 
The scheduler can then query the model with possible resource allocations for each microservice, and select the one that meets QoS 
with the least necessary resources.}

A straightforward way to achieve this is designing an ML model that predicts the immediate end-to-end tail latency 
as a function of resource allocations and utilization, since QoS is defined in terms of latency, and comparing the predicted latency to 
the measured latency during deployment is straightforward. The caveat of this approach is the delayed queueing effect described in Sec.~\ref{sec:challenges}, 
whereby the impact of an allocation decision would only show up in performance later. 
As a resolution, we experimented with training a neural network (NN) to predict latency distributions over a future time window: % in the future; 
for example, the latency for each second over the next five seconds. However, we found that the prediction accuracy rapidly decreased the further into 
the future the NN tried to predict, as predictions were based only on the collected current and past metrics (resource utilization and latency), 
which were accurate enough for immediate-future predictions, but were insufficient to capture how dependencies between microservices would cause performance 
to evolve later on. %further into the future. 
%is a sound input for predicting latency of next cycle, however such information does not exist for predicting further future.

%BT works well because as long as you get the prediction window right, you can use the current state to estimate the latency once the inertia passes, you don't need the intermediate states, because they'll only affect the system later on. Then why doesn't the CNN also predict accurately? Because it tries to predict every cycle instead of once into the further future... 

Considering the difficulty of predicting latency further into the future, we set an alternative goal: predict the latency 
of the immediate future, such that imminent QoS violations are identified quickly, but only predict the probability of experiencing a QoS violation later on, instead of the exact latency of each decision interval. This binary classification is a much more contained 
problem than detailed latency prediction, and still conveys enough information to the resource manager on performance events, e.g., QoS violations, 
that may require immediate action in the present. 

\begin{figure}[!h]
% \vspace{-0.18in}
    \centering
	   % \vspace{-0.16in}
    %\small (a) Hotel Reservation under constant and (b) diurnal load.\\
    \begin{tabular}{@{}c@{}}
    \includegraphics[scale=0.25,viewport=220 30 440 470]{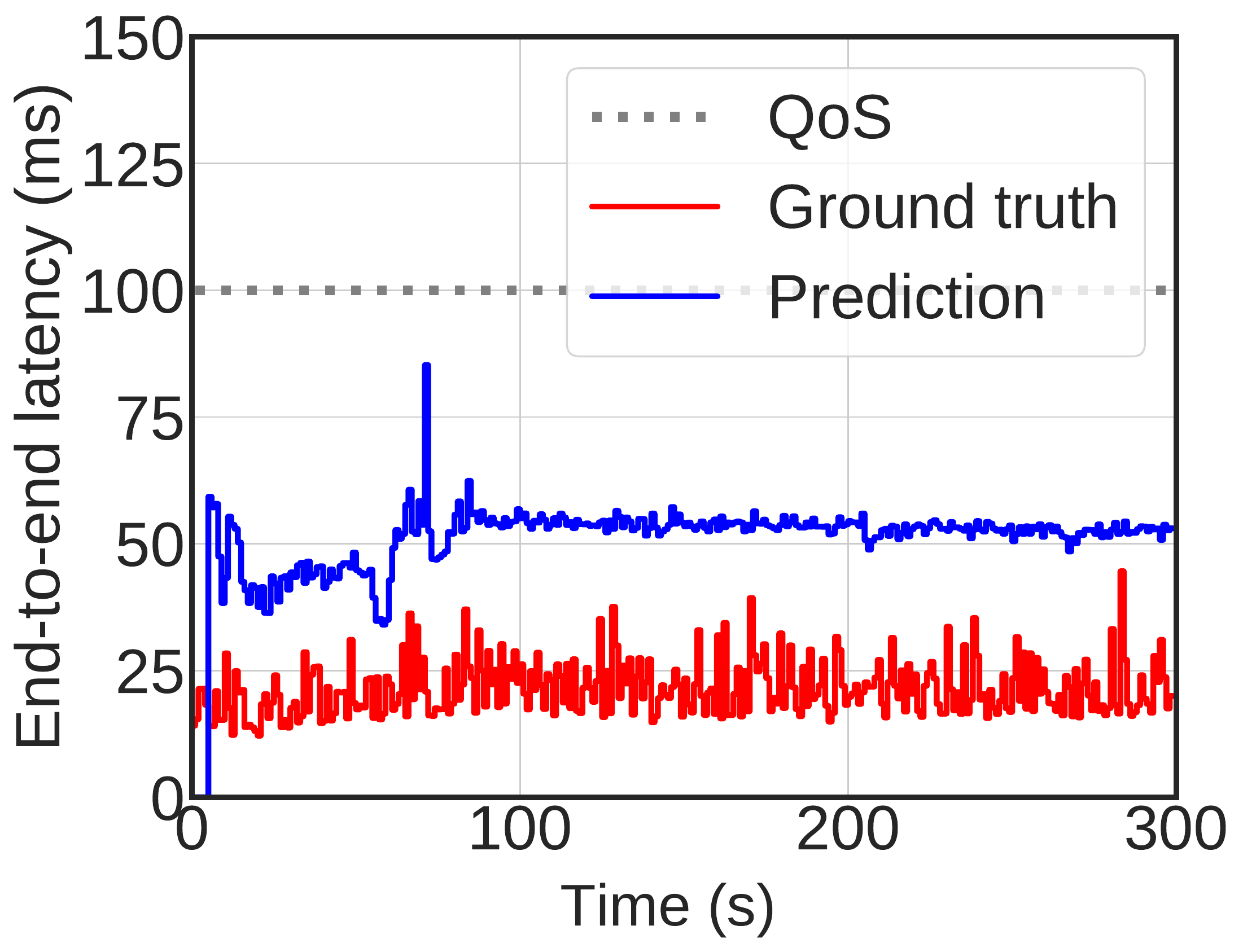}
    \end{tabular}\\
    % \vspace{-0.08in}
	\caption{\label{fig:overpredict} Multi-task NN overpredicts Social Network latency,  {\color{black} due to the semantic gap between the QoS violation probability, a value between 0 and 1, and the latency, a value that is not strictly bounded. }}
% \vspace{-0.08in}
\end{figure} 
% {\noindent and it tries to infer quantities with very different value ranges -- latencies and violation probabilities. }%which hurts performance and efficiency. 
%For the immediate future, we still want to get a latency prediction because first, the resource manager can use predicted latency to improve system performance, and secondly, as mentioned previously, predicted latency is an intuitive indication of the quality of the model. In online deployment, the model can be distrusted and pushed offline if the predicted latency continue to differ from ground truth by non-trivial margin.

An intuitive method for this are multi-task learning NNs that predict the latency of the next interval, and the QoS violation 
probability in the next few intervals. However, the multi-task NN considerably overpredicts tail latency 
and QoS violation probability, as shown in Fig.~\ref{fig:overpredict}. 
Note that the gap between prediction and ground truth does not indicate a constant difference, 
which could be easily learned by NNs with strong overfitting capabilities. We attribute the overestimation to interference 
caused by the semantic gap between the QoS violation probability, a value between 0 and 1, and the latency, a value that is not strictly bounded. %can be as high as a few hundred.

%The dual-task neural network optimizes a linear regression problem for predicting the latency and a logistic regression problem as predicting the probability of QoS violation jointly.
%Based on our empirically evaluation, the dual-task model performs reasonably on the training and validation datasets.
% \begin{wrapfigure}[6]{r}{0.30\textwidth}
%     %\vspace{-0.18in}
%     \centering
% 	\vspace{-0.26in}
% 	\hspace*{-0.2in}\includegraphics[scale=0.25,trim=0 0 0 9.6cm, clip=true]{figures/Sinan_model.pdf}
% 	\vspace{-0.24in}
%     \caption{Sinan's two-stage model. }
%     \label{fig:model_diagram}
% \end{wrapfigure}
% {\noindent and it tries to infer quantities with very different value ranges -- latencies and violation probabilities. }%which hurts performance and efficiency. 

To address this, we designed a two-stage model: first, a CNN that predicts the end-to-end latency of the next timestep with high accuracy, 
and, second, a Boosted Trees (BT) model that estimates the probability for QoS violations further into the future, using the latent variable extracted by CNN. 
BT is generally less prone to overfitting than CNNs, since it has much fewer tunable hyperparameters than NNs; mainly the number of trees and tree depth. By using two separate models, 
Sinan is able to optimize each model for the respective objective, and avoid the overprediction issue of using a joint, expensive model for both tasks. 
We refer to the CNN model as the \textit{short-term latency predictor}, and the BT model as the \textit{long-term violation predictor}. 

%The decision window of the BT is sized such that it accounts for the system's inertia in reacting to resource allocation decisions (delayed queueing effect), 
%therefore it is not contingent on accurately predicting the service's latency on each interval until then. If it has inertia of 5 intervals, I just need to know the current 
%state to predict t+5, while the latency model tried to predict, e.g., t+2 using t and t+1, which may not be accurate, because you don't know how t+1 will affect t+2. 
%By predicting directly at the end of the inertia window, you know that only the current state and nothing between now and then will affect that future state. 

\subsection{Latency Predictor}

{\color{black} As discussed in Section~\ref{sec:challenges}, the CNN needs to account for both the dependencies across microservices, 
and the timeseries pattern of resource usage and application performance. Thus, both the application topology and the timeseries information are encoded in the input of the CNN. 
The input of the CNN includes the following three parts: 
\begin{enumerate}
\item an ``image'' (3D tensor) consisting of per-tier resource utilization within a past time window. The y-axis of the ``image'' corresponds to different microservices, with consecutive tiers in adjacent rows, the x-axis corresponds to the timeseries, with one timestep per column, and the z-axis (channels) corresponds to resource metrics of different tiers, including CPU usage, memory usage (resident set size and cache memory size) and network usage (number of received and sent packets), which are all retrieved from Docker’s cgroup interface. Per-request tracing is not required.
\item a matrix of the end-to-end latency distribution within the past time window, and 
\item the examined resource configuration for the next timestep, which is also encoded as a matrix. 
\end{enumerate}
In each convolutional (Conv) layer of the CNN,
a convolutional kernel ($k\times k$ window) processes information of $k$ adjacent tiers within a time window containing $k$ timestamps. 
The first few Conv layers in the CNN can thus infer the dependencies of their adjacent tiers over a short time window, and later layers 
observe the entire graph, and learn interactions across all tiers within the entire time window of interest. The latent representations 
derived by the convolution layers are then post-processed together with the latency and resource configuration information, 
through concatenation and fully-connected (FC) layers to derive the latency predictions. In the remainder of this section, 
we first discuss the details of the network architecture, and then introduce a custom loss function 
that improves the prediction accuracy by focusing on the most important latency range. }

\begin{figure}[t]
    \centering 
    \includegraphics[scale=0.14]{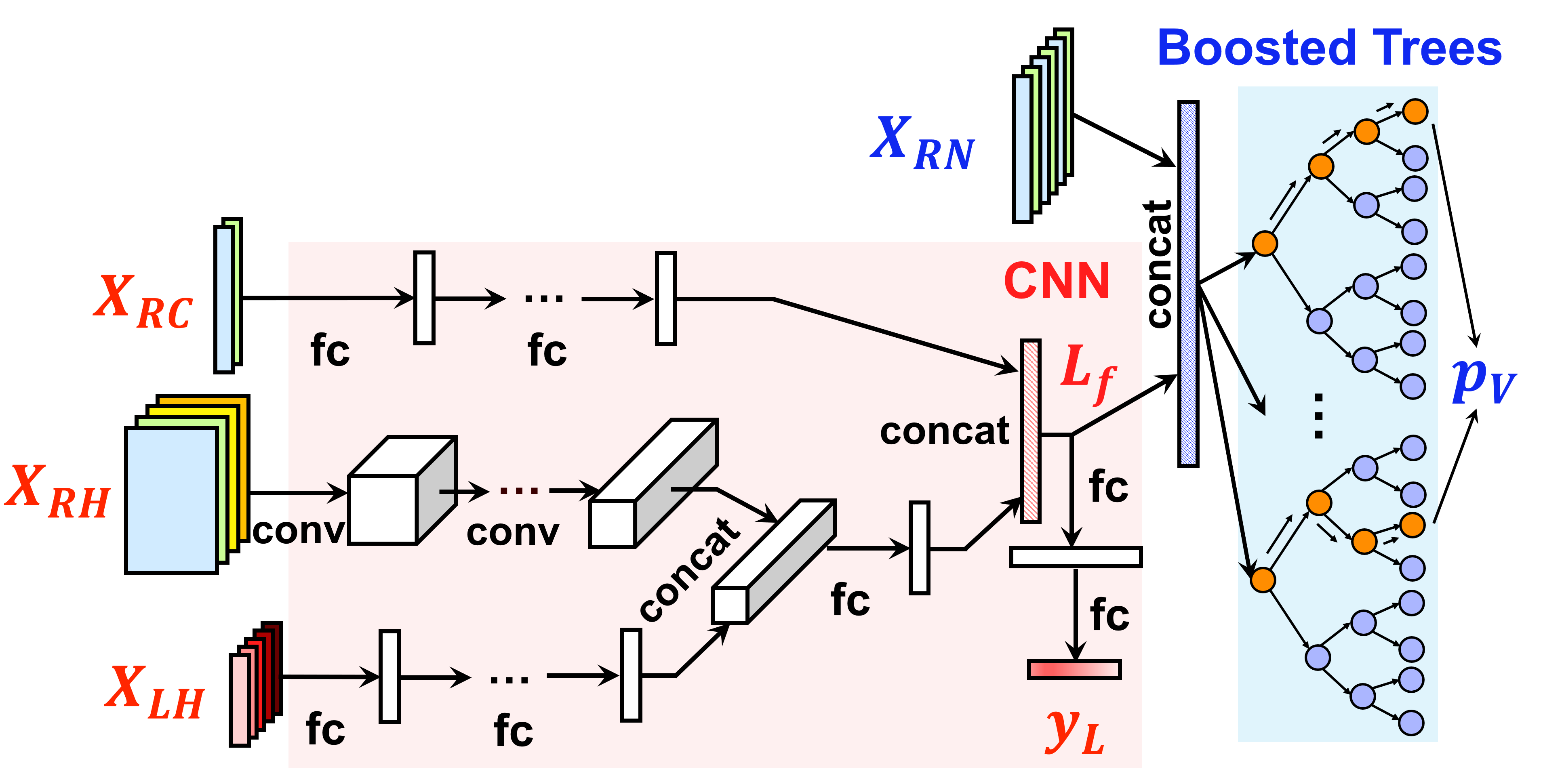}
	\caption{Sinan's hybrid model, consisting of a CNN and a Boosted Trees (BT) model. The CNN extracts the latent variable ($L_f$) 
    and predicts the end-to-end latency ($y_L$). The BT take the latent variable and proposed resource allocation, 
    and predicts the probability of a QoS violation ($p_V$).}
    \label{fig:models}
    % \vspace{-0.15in}
\end{figure}

As shown in Fig.~\ref{fig:models}, the latency predictor takes as input the resource usage history ($X_{RH}$), 
the latency history ($X_{LH}$), and the resource allocation under consideration for the next timestep ($X_{RC}$), 
and predicts the end-to-end tail latencies ($y_L$) (95$^{th}$ to 99$^{th}$ percentiles) of the next timestep.

$X_{RH}$ is a 3D tensor whose x-axis is the $N$ tiers in the microservices graph, 
the y-axis is $T$ timestamps ($T>1$ accounts for the non-Markovian nature of microservice graph), 
and channels are $F$ resource usage information related to CPU and memory. The set of necessary and sufficient resource metrics 
is narrowed down via feature selection. $X_{RC}$ and $X_{LH}$ are 2D matrices. For $X_{RC}$, the x-axis is the $N$ tiers and the y-axis 
the CPU limit. For $X_{RH}$, the x-axis is $T$ timestamps, and the y-axis are vectors of different latency percentiles (95$^{th}$ to 99$^{th}$). 
The three inputs are individually processed with Conv and FC layers, 
and then concatenated to form the latent representation $L_{f}$, from which the predicted tail latencies $L_{f}$ are derived with another FC layer.

%\textbf{Custom Loss Function} \hspace{3pt}
The CNN minimizes the difference between predicted and actual latency, using the squared loss function below:
% \vspace{-0.08in}
\begin{equation}
    \mathcal{L}(X, \hat{y}, W) = \sum_i^n (\hat{y_i}-f_W(x_i))^2
    \label{eq:square_loss}
\end{equation}
% \vspace{-0.04in}
where $f_W(\cdot)$ represents the forward function of the CNN, $\hat{y}$ is the ground truth, 
and $n$ is the number of training samples. Given the spiking behavior of interactive microservices 
that leads to very high latency, the squared loss in Eq.~\ref{eq:square_loss} tends to overfit for training samples 
with large end-to-end latency, leading to latency overestimation in deployment. 
%It is worth mentioning that this technique only mitigates overfitting of predicted latency 
%for the next interval, and does not improve predictions further into the future, as described above.
% of further future obviously, due to the reasons described before.
Since the latency predictor aims to find the best resource allocation within a tail latency QoS target, 
the loss should be biased towards training samples whose end-to-end latencies are $\leq QoS$. 
Therefore, we use a scaling function to scale both the predicted and actual end-to-end latency 
before applying the squared loss function. The scaling function ($\phi(\cdot)$) is:
% \vspace{-0.08in}
\begin{equation}
\phi(x)=
\begin{cases}
    x& \text{$x$} \leq \text{$t$}\\
    t + \frac{x-t}{1+\alpha(x-t)} & \text{$x > t$}
\end{cases}
% \vspace{-0.08in}
\label{eq:1}
\end{equation}
% \vspace{-0.02in}
where the latency range is $(0,t)$, and the hyper-parameter $\alpha$ can be tuned for different decay effects. 
%The scaling factor helps generalize the model to non-problematic cases, when the system is not saturated to avoid biases from the few QoS violations. 
Fig.~\ref{fig:loss_func} shows the scaling function with $t = 100$ and $\alpha = 0.005, 0.01, 0.02$. %, respectively.  
It is worth mentioning that scaling end-to-end latencies 
only mitigates overfitting of the predicted latency for the next decision interval, and does not improve predictions further into the future, as described above.
We implement all CNN models using MxNet~\cite{mxnet}, and trained them with Stochastic Gradient Descent (SGD). 
%a high performance deep learning framework. All models are trained using Stochastic Gradient Descent (SGD).
%\vspace{0.04in}

% \vspace{-0.08in}
\subsection{Violation Predictor}
% \vspace{-0.06in}

The violation predictor addresses the binary classification task of predicting whether a given allocation 
will cause a QoS violation further in the future, to filter out undesirable actions. %resource options.
Ensemble methods are good candidates as they are less prone to overfitting. We use Boosted Trees~\cite{boosting}, 
which realizes an accurate non-linear model by combining a series of simple regression trees. 
It models the target as the sum of trees, each of which maps features to a score. 
The final prediction is determined by accumulating scores across all trees.

\begin{minipage}{\linewidth}
  \begin{minipage}{0.51\linewidth}
  %\vspace{0.08in}
    %\centering
	  %\begin{table}
%\begin{adjustbox}{width=1.0\linewidth}
    \begin{tabular}{cp{3.6cm}}
    \footnotesize
%\toprule
    \textbf{{\footnotesize {Param}}} & \textbf{\footnotesize{{Definition}}}\\
%\toprule
    \textbf{\footnotesize{{k}}} & {\footnotesize{{future timesteps in BT}}} \\[0.06cm]
    \textbf{\footnotesize{{T}}} & {\footnotesize{{past timesteps in CNN\&BT}}} \\[0.06cm]
    \textbf{\footnotesize{{N}}} & {\footnotesize{{application tiers}}} \\[0.06cm]
    \textbf{\footnotesize{{M}}} & {\footnotesize{{latency percentiles}}} \\[0.06cm]
    \textbf{\footnotesize{{F}}} & {\footnotesize{{resource statistics}}} \\[0.06cm]
    \textbf{\footnotesize{{R}}} & {\footnotesize{{allocated resources}}} \\[0.06cm]
%\bottomrule
\end{tabular}
% 	\vspace{-0.05in}
      %\caption{Parameters for the hybrid ML model.}
\captionof{table}{{ML model parameters. }}
      %\end{table}
      %\end{adjustbox}
    \end{minipage}
    \hspace{0.1cm}
  \begin{minipage}{0.41\linewidth}
    \centering
    %\vspace{-0.12in}
    \includegraphics[scale=0.275,viewport=30 0 400 260]{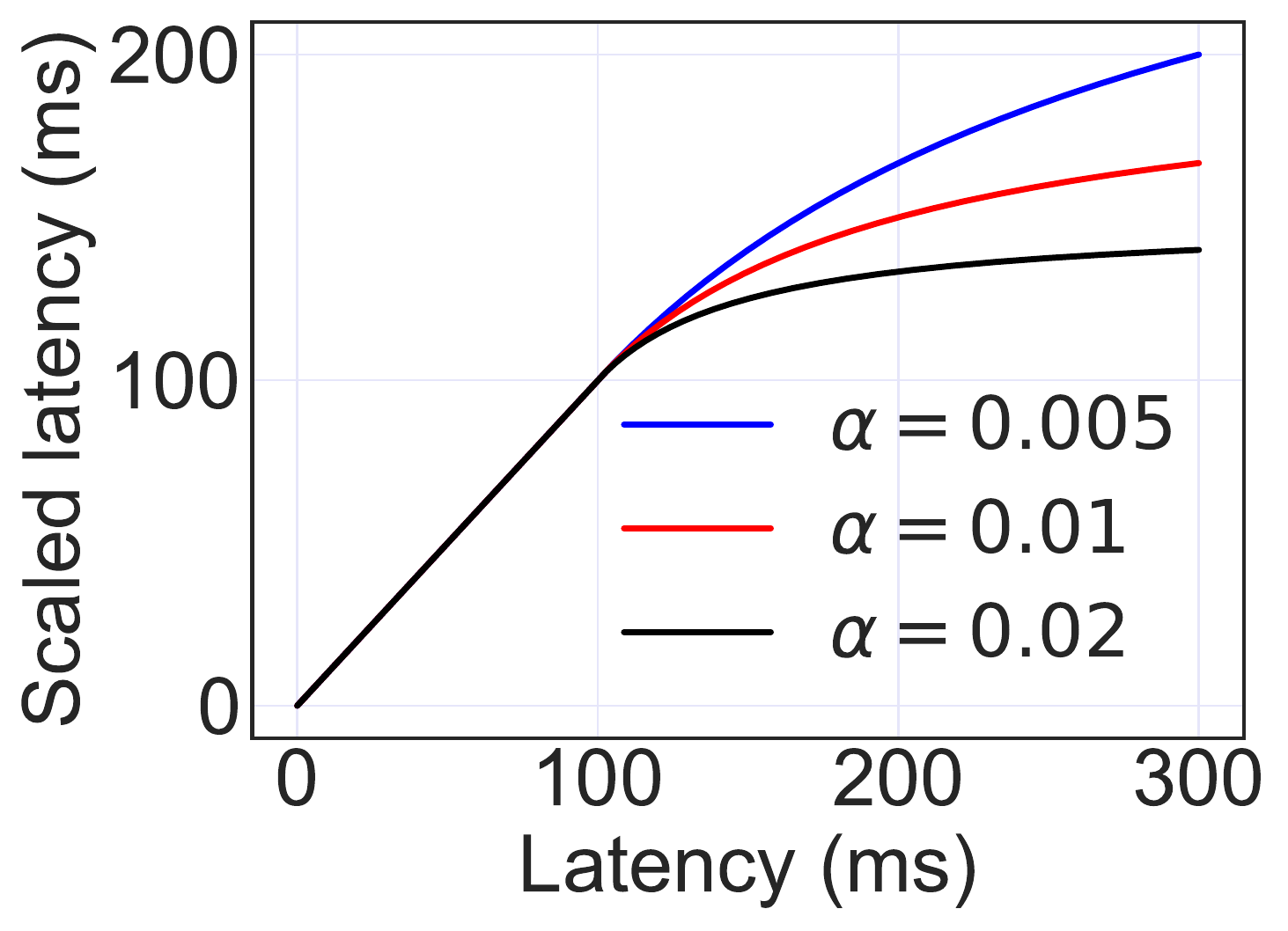}
    % \vspace{-0.25cm}
    \captionof{figure}{\label{fig:loss_func}Scale function $\phi(\cdot)$ with different $k$.}
  \end{minipage}
  \end{minipage}

To further reduce the computational cost and memory footprint of Boosted Trees, 
we reuse the compact latent variable $L_f$ extracted from the CNN as its input. 
Moreover, since the latent variable $L_f$ is significantly 
smaller than $X_{RC}$, $X_{RH}$, and $X_{LH}$ in dimensionality, using $L_f$ as 
the input also makes the model more resistant to overfitting. 

Boosted Trees also takes resource allocations as input. During inference, we simply use 
the same resource configuration for the next $k$ timesteps to predict whether it will 
cause a QoS violation $k$ steps in the future. As shown in Fig.~\ref{fig:models}, 
each tree leaf represents either a violation or a non-violation with a continuous score. 
For a given example, we sum the scores for all chosen violation ($s_{V}$) and non-violation 
leaves ($s_{V}$) from each tree. The output of BT is the predicted probability of QoS violation ($p_{V}$), 
which can be calculated as $p_{V} = \frac{e^{s_{V}}}{e^{s_{V}}+e^{s_{NV}}}$. For the violation predictor  
we leverage XGBoost~\cite{xgboost}, a gradient tree boosting framework that improves scalability using sparsity-aware 
approximate split finding. %, to build and train our model. 

% It is worth noting that the latency and violation predictors are trained separately without interference. 
We first train the CNN and then BT using the extracted latent variable from the CNN. 
The CNN parameters (number of layers, channels per layer, weight decay etc.) 
and XGBoost (max tree depth) are selected based on the validation accuracy. 

% \vspace{-0.06in}

% \newcommand{\pluseq}{\mathrel{+}=}

\section{System Design}
\label{sec:design}

We first introduce Sinan's overall architecture, and then discuss the data collection process, 
which is crucial to the effectiveness of the ML models, and Sinan's online scheduler. %ing algorithm. % based on aforementioned ML models.

% \note{This is missing a figure that shows the design on Sinan. Added the previous below, update as needed. }

\subsection{System Architecture}
% \vspace{-0.06in}
% \subsubsection{Overview}
\label{sec:design_overview}

\begin{figure}
\centering
  \includegraphics[scale=0.16,viewport=450 50 1000 500]{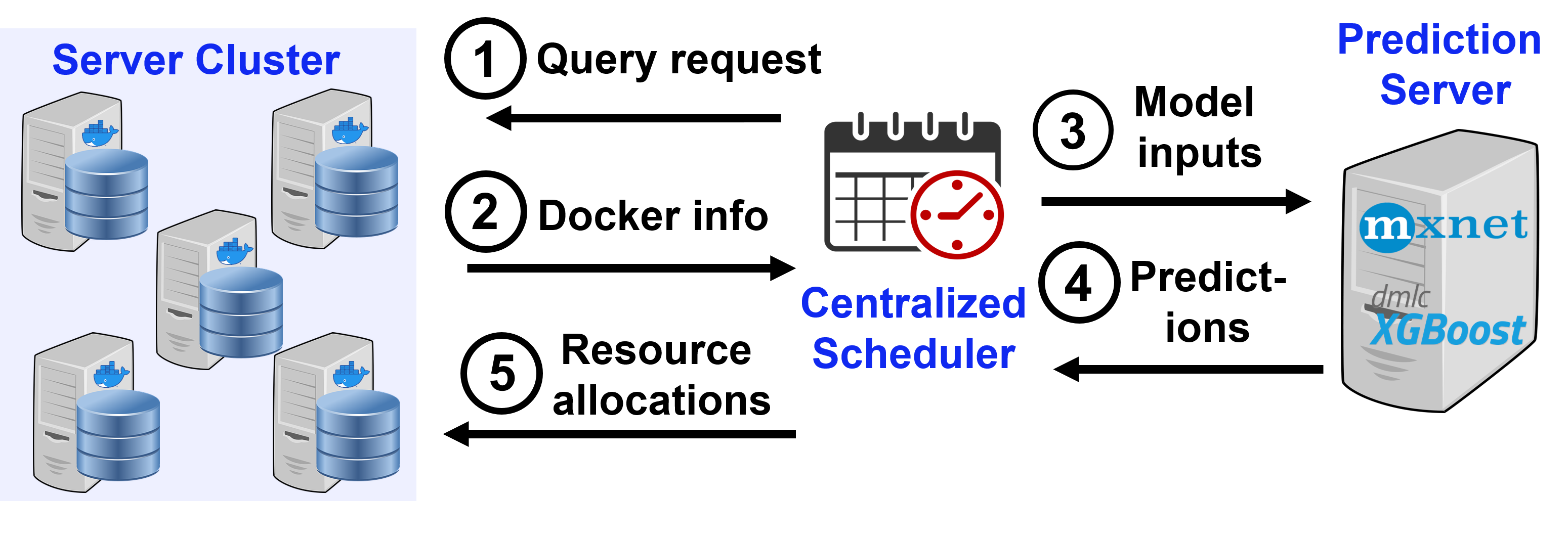}
	\caption{\textcolor{black}{Sinan's system architecture. As user requests are being received, Sinan collects resource and performance metrics through Docker and Jaeger, 
	inputs the collected metrics to the ML models, and uses the models' output to accordingly allocate resources for each tier. Allocation decisions are re-evaluated periodically online. }}
  \label{fig:sys}
\end{figure}

Sinan consists of three components: a centralized scheduler, distributed operators deployed on each server/VM, 
and a prediction service that hosts the ML models. Figure~\ref{fig:sys} shows an overview of Sinan's architecture. 
% , as shown in Fig.~\ref{fig:sys}. 

Sinan makes decisions periodically. In each 1s decision interval (consistent with the granularity at which QoS is defined), 
the centralized scheduler queries the distributed operators to obtain the CPU, memory, and network utilization 
of each tier in the previous interval. Resource usage is obtained from Docker's monitoring infrastructure, 
and only involves a few file reads, incurring negligible overheads. 
Aside from per-tier information, the scheduler also queries the API gateway to get user load statistics from the workload generator. 
%implemented via the workload generator, for simplicity, in our setting. 
The scheduler %receives this data, 
sends this data to the hybrid ML model, which is responsible for evaluating the impact of different resource allocations. % decisions. 
Resource usage across replicas of the same tier are averaged before being used as inputs to the models.
%Decisions are limited to a predefined set of operations to minimize scheduling overheads. %enable timely evaluation and speed up training data collection.
Based on the model's output, Sinan chooses an allocation vector that meets QoS using 
the least necessary resources, and communicates its decision to the per-node agents for enforcement.

Sinan focuses on compute resources, which are most impactful to microservice performance. 
Sinan explores sub-core allocations in addition to allocating multiple cores per microservice to 
avoid resource inefficiencies for non-resource demanding tiers, and enable denser colocation. 
%Sub-core CPU allocations are used instead of frequency management power management We choose not to perform power management because power management is limited to platform and hardware and thus lacks generality. For example, power management can not be carried out on public cloud.

%\subsubsection{Implementation}

% \vspace{-0.06in}
\subsection{Resource Allocation Space Exploration}
% \vspace{-0.06in}
\label{sec:data_collection}

Representative training data is key to the accuracy of any ML model. Ideally, test data encountered during online deployment 
should follow the same distribution as the training dataset, so that covariate shift is avoided. Specifically for our problem, 
the training dataset needs to cover a sufficient spectrum of application behaviors that are likely to occur during online deployment. 
Because Sinan tries to meet QoS without sacrificing resource efficiency, it must efficiently explore the boundary of the resource allocation space, 
where points using the minimum amount of resources under QoS reside. 
We design the data collection algorithm as a multi-armed bandit process~\cite{gittins2011multi}, where each tier is an independent arm, 
with the goal of maximizing the knowledge of the relationship between resources and end-to-end QoS.

The data collection algorithm approximates the running state of the application with a tuple $(rps,~lat_\text{cur},~lat_\text{diff})$, where $rps$ 
is the input requests per second, $lat_\text{cur}$ is the current tail latency, 
and $lat_\text{diff}$ is the tail latency difference from the previous interval, to capture the rate of consuming or accumulating queues. 
Every tier is considered as an arm that can be played independently, 
by adjusting its allocated resources. For each tier, we approximate the mapping 
between its resources and the end-to-end QoS as a Bernoulli distribution, with probability $p$ of meeting 
the end-to-end QoS, and we define our information gain from assigning certain amount of resources to a tier, 
as the expected reduction of confidence interval of $p$ for the corresponding Bernoulli distribution. 
At each step for every tier, we select the operation that maximizes the information gain, as shown in Eq.~\ref{eq:mab_op}, 
where $op_T^s$ is an action selected for tier $T$ at running state $s$, $n$ are the samples 
collected for the resulting resource assignment after applying $op$ on tier $T$ at state $s$, $p$ is the previously-estimated probability of meeting QoS, 
and $p_+$ and $p_-$ are the newly-estimated probabilities of meeting QoS, when the new sample meets or violates QoS respectively. 
% The $p$ and $n_{Top}^s$ are initialized as 0.5 and 0.1 in our experiments. 
Each operation's score is multiplied by a predefined coefficient $C_{op}$ to encourage meeting QoS and reducing %resource 
overprovisioning. 

% \vspace{-0.24in}
\begin{equation}
\small
\begin{gathered}
    % \mathcal{L}(X, \hat{y}, W) = \sum_i^n (\hat{y_i}-f_W(x_i))^2
    op_T^s=\text{arg}\max_{op} C_{op}\cdot(\sqrt{\frac{p(1-p)}{n}}-p\sqrt{\frac{p_+(1-p_+)}{n+1}} \\
    -(1-p)\sqrt{\frac{p_-(1-p_-)}{n+1}})
% \vspace{-0.05in}
\label{eq:mab_op}
\end{gathered}
\end{equation}
% \vspace{-0.04in}

% \vspace{-0.24in}
% \begin{multline}
%     % \mathcal{L}(X, \hat{y}, W) = \sum_i^n (\hat{y_i}-f_W(x_i))^2
%     op_T^s=\text{arg}\max_{op} (\sqrt{\frac{p(1-p)}{n}}-p\sqrt{\frac{p_+(1-p_+)}{n+1}}\\
%     -(1-p)\sqrt{\frac{p_-(1-p_-)}{n+1}})\cdot C_{op}
% \vspace{-0.36in}
%     \label{eq:mab_op}
% \end{multline}
% \vspace{-0.04in}

By choosing operations that maximize Equation.~\ref{eq:mab_op}, the data collection algorithm is incentivized to explore the 
boundary points that meet QoS with the minimum resource amount, since exploring allocations that definitely meet or violate QoS 
(with $p = 1$ or $p = 0$) has at most 0 information gain. Instead, the algorithm prioritizes exploring 
resource allocations whose impact on QoS is nondeterministic, like those with $p=0.5$. It is also worth noting that the state encoding 
and information gain definition are simplified approximations of the actual system, with the sole purpose of containing the exploration 
process in the region of interest. Eventually, we rely on ML to extract the state representation that incorporates inter-tier dependencies in the microservice graph.

To prune the action space, Sinan enforces a few rules on both data collection and online scheduling. 
First, the scheduler is only allowed to select out of a predefined set of operations. Specifically 
in our setting, the operations include reducing or increasing the CPU allocation by 0.2 up to 1.0 CPU, 
and increasing or reducing the total CPU allocation of a service by 10\% or 30\%. These ratios are selected 
according to the AWS step scaling tutorial~\cite{aws_step_scaling}; as long as the granularity of CPU allocations does not change, other resource ratios also work without retraining the model. 
Second, an upper limit on CPU utilization is enforced on each tier, to avoid overly aggressive resource downsizing that can lead to long queues and dropped requests. 
Third, when end-to-end tail latency exceeds the expected value, Sinan disables resource reclamations so that the system can recover as soon as possible. 
A subtle difference from online deployment is that the data collection algorithm explores resource allocations in the $[0, QoS + \alpha]$ tail latency region, 
where $\alpha$ is a small value compared to QoS. The extra $\alpha$ allows the data collection process to explore allocations 
that cause slight QoS violations without the pressure of reverting to states that meet QoS immediately, such that the ML models 
are aware of boundary cases, and avoid them in deployment. In our setting $\alpha$ is 20\% of QoS empirically, to adequately explore 
the allocation space, without causing the tail latency distribution to deviate too much from values that would be seen in deployment. 
Collecting data exclusively when the system operates nominally, or randomly exploring the allocation space does not fulfill these requirements. 

\begin{figure}[h!]
\centering
\begin{minipage}{.5\linewidth}
  \centering
  \includegraphics[width=1.0\linewidth]{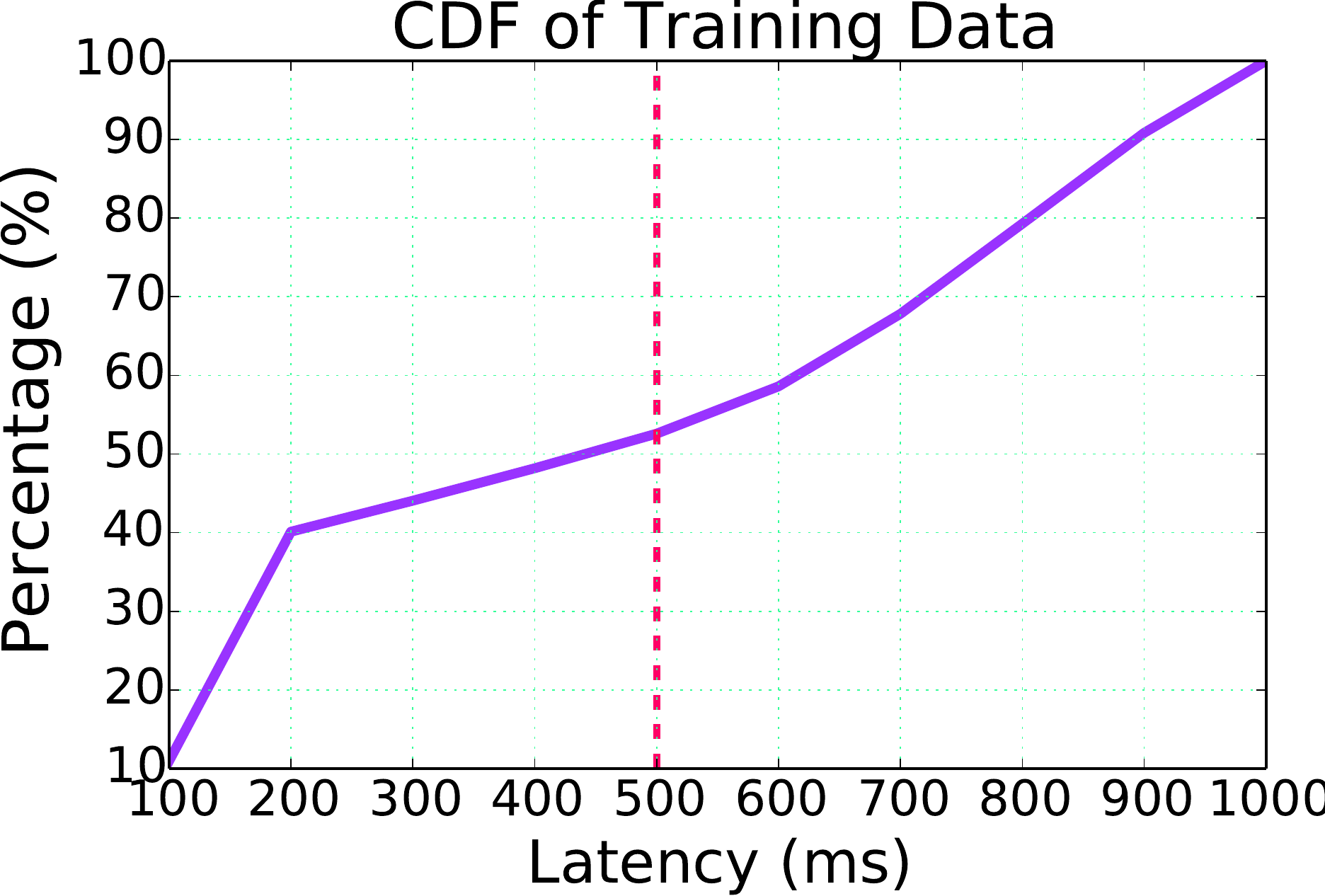}\\
%   \vspace{-0.08in}
  %\small (a) Random data collection.
\label{fig:cnn_err}
\end{minipage}%
\begin{minipage}{.5\linewidth}
  \centering
  \includegraphics[width=1.15\linewidth]{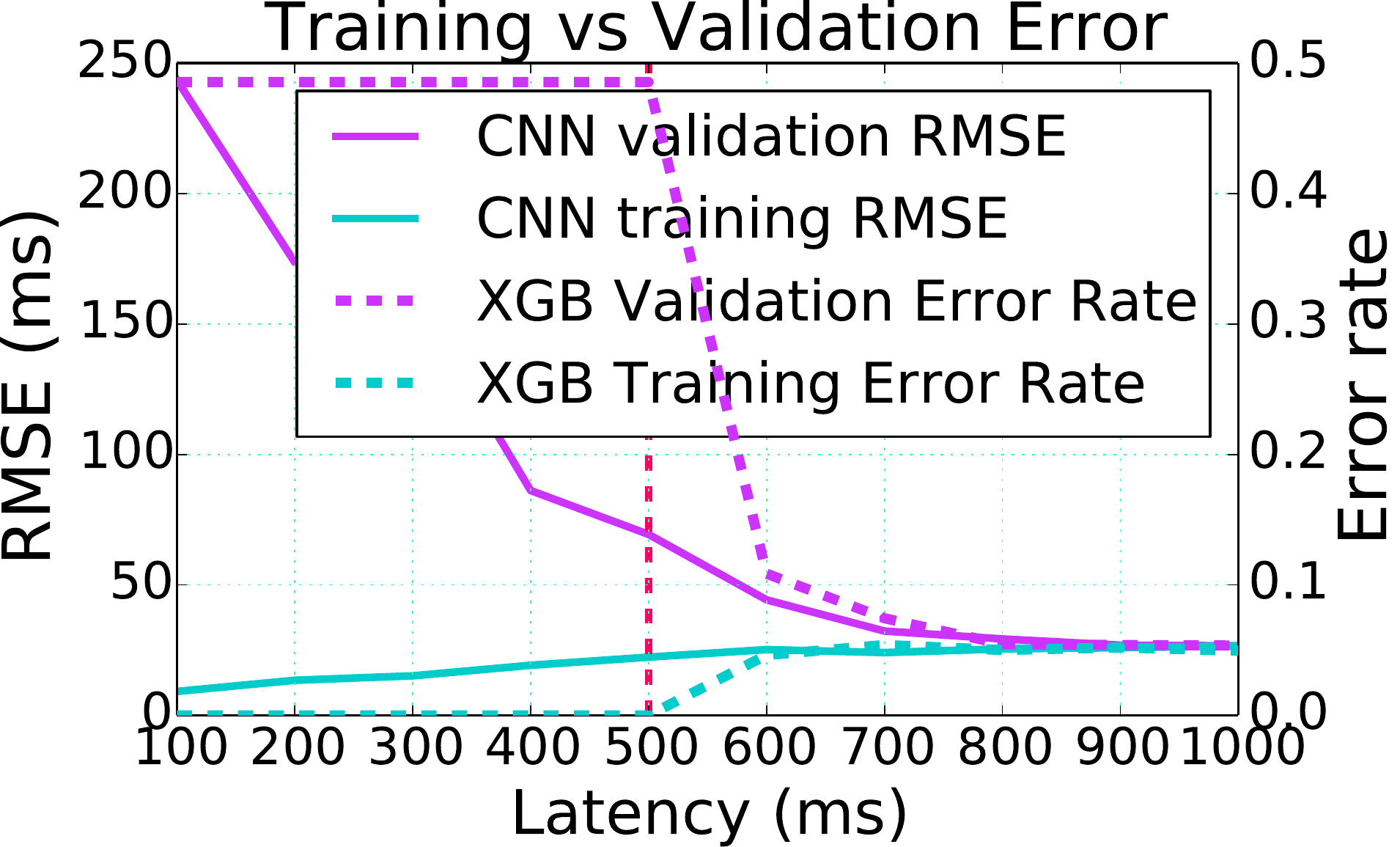}\\
%   \vspace{-0.08in}
  %\small (b) Autoscaling data collection.
\label{fig:xgb_err}
\end{minipage}\caption{Training dataset latency distribution and ML training vs. validation error with respect to dataset latency range. {\color{black} The training dataset includes an approximately balanced set of samples between those that preserve and those that violate QoS. If the training dataset does not 
include any samples that violate QoS (500ms), both the CNN and XGBoost experience serious overfitting, greatly mispredicting latencies and QoS violations. }}
\label{fig:ml_lat_sens}
%   \vspace{-0.08in}
\end{figure}

Fig.~\ref{fig:ml_lat_sens} shows the latency distribution in the training dataset, and how the training and validation error 
of the model changes with respect to the latency range observed in the training dataset, for the Social Network application. 
In the second figure, the x-axis is the latency of samples in the training dataset, the left y-axis is the root mean squared error 
RMSE of the CNN, and the right y-axis represents the classification error rate of XGBoost. Each point's y-axis value is the model's training 
and validation error when trained only with data whose latency is smaller than the corresponding x-value. If the training dataset does not 
include any samples that violate QoS (500ms), both the CNN and XGBoost experience serious overfitting, greatly mispredicting latencies and QoS violations. 

\begin{figure}[h!]
\centering
%   \vspace{-0.08in}
\begin{minipage}{.5\linewidth}
  \centering
  \includegraphics[width=1.05\linewidth]{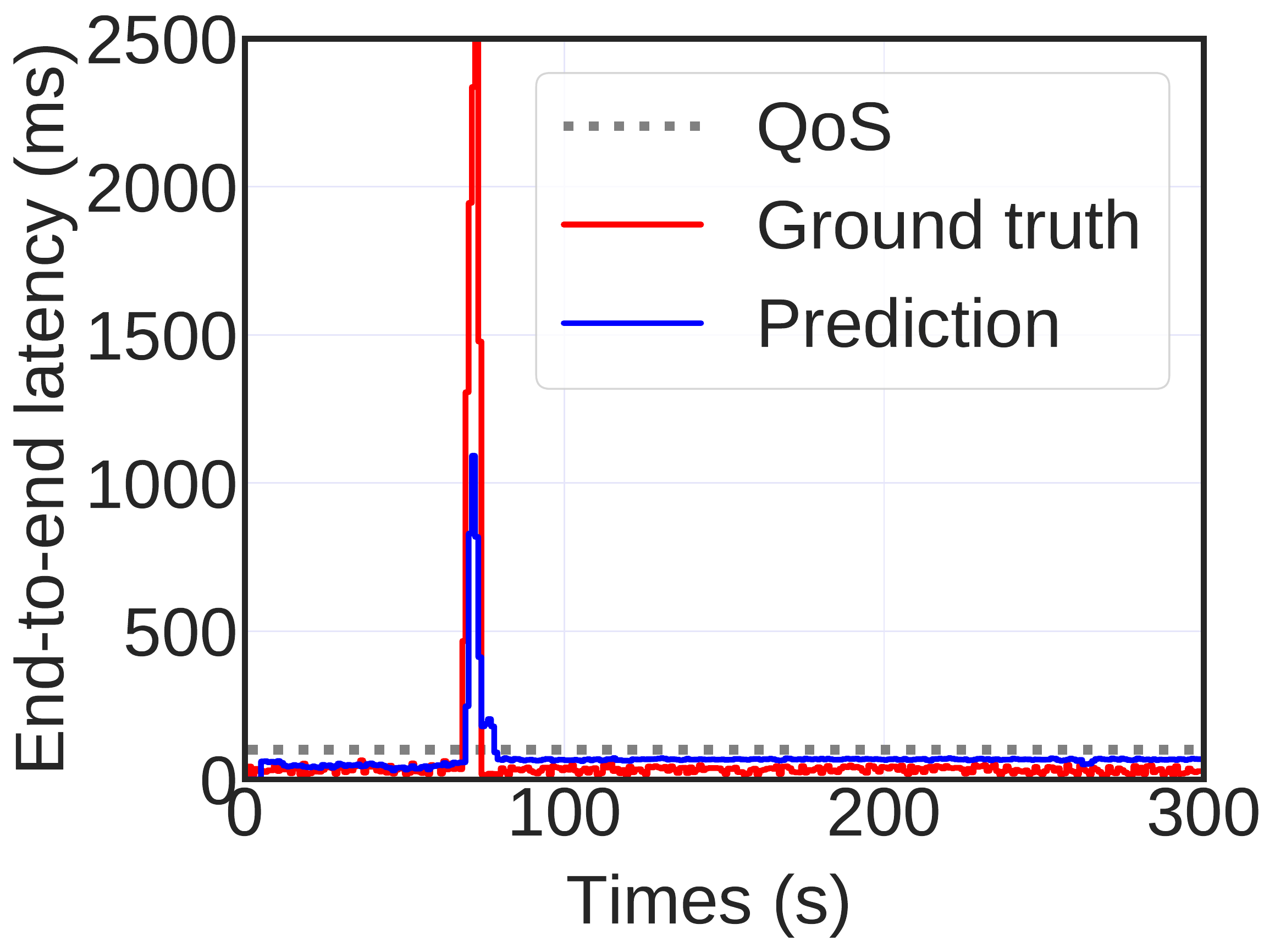}\\
%   \vspace{-0.08in}
  \small (a) Autoscaling data collection.
\label{fig:random_data}
\end{minipage}%
\begin{minipage}{.5\linewidth}
  \centering
  \includegraphics[width=1.05\linewidth]{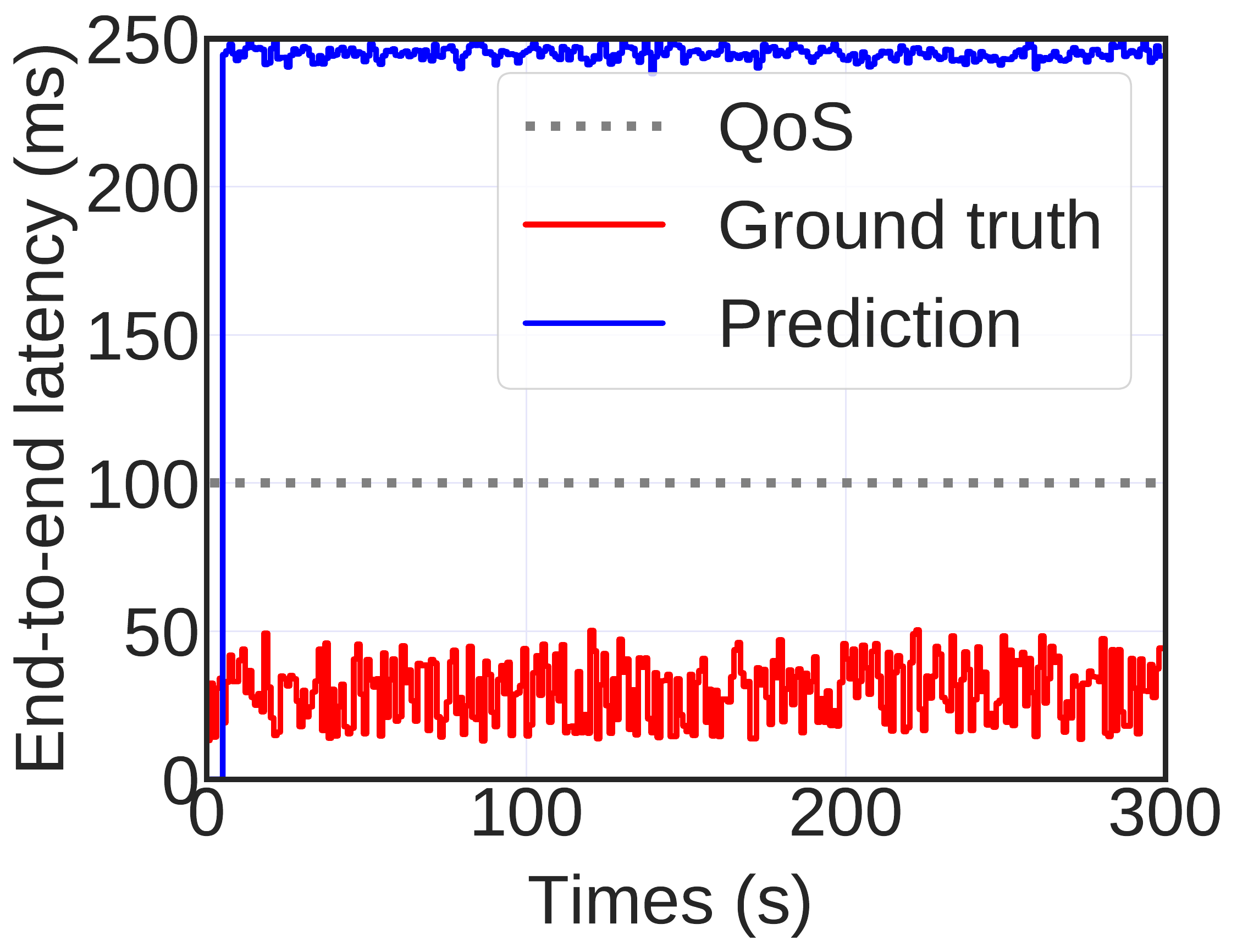}\\
%   \vspace{-0.08in}
  \small (b) Random data collection.
\label{fig:as_data}
\end{minipage}\caption{Comparison of predicted and true latency with (a) autoscaling and (b) random data collection schemes. {\color{black} When using autoscaling, the model significantly underestimates latency due to insufficient training samples of QoS violations, and causes 
large spikes in tail latency, forcing the scheduler to use all available resources to prevent further violations. 
On the other hand, when the model is trained using random profiling, 
it constantly overestimates latency and prohibits any resource reduction, leading to resource overprovisioning. } }
\label{fig:data_collection_comparison}
%   \vspace{-0.12in}
\end{figure}

Fig.~\ref{fig:data_collection_comparison} shows data collected using data collection mechanisms that do not curate the dataset's distribution. 
Specifically, we show the prediction accuracy when the training dataset is collected when autoscaling is in place (a common resource management scheme in most clouds), 
and when resource allocations are explored randomly. 
As expected, when using autoscaling, the model does not see enough cases that violate QoS, and hence seriously underestimates latency and causes 
large spikes in tail latency, forcing the scheduler to use all available resources to prevent further violations. 
On the other hand, when the model is trained using random profiling, 
it constantly overestimates latency and prohibits any resource reduction, highlighting the importance of jointly designing the data collection algorithms and the ML models. 

\noindent{{\bf{{Incremental and Transfer Learning:}}}{ Incremental retraining can be applied to accommodate changes to the deployment strategy or microservice updates. 
In cases where the topology of the microservice graph is not impacted, such as hardware updates and change of public cloud provider, transfer learning techniques such as fine tune can be used to train the ML models in the background with newly collected data. 
If the topology is changed, the CNN needs to be modified to account for removed and newly-added tiers. }} 

% \vspace{-0.14in}
\noindent{\bf{{Additional resources: }}}Sinan can be extended to other system resources. Several resources, such as network bandwidth and memory capacity act like thresholds, below 
which performance degrades dramatically, e.g., network bandwidth~\cite{chen2019parties}, or the application experiences out of memory errors, and 
can be managed with much simpler models, like setting fixed thresholds for memory usage, or scaling proportionally with respect to user load for network bandwidth. 

\subsection{Online Scheduler} 
\label{sec:scheduler}
%\vspace{-0.1in}

During deployment, the scheduler evaluates resource allocations using the ML models, 
and selects appropriate allocations that meet the end-to-end QoS without overprovisioning. 

Evaluating all potential resource allocations online would be prohibitively expensive, especially for complex microservice topologies. 
Instead, the scheduler evaluates a subset of allocations following the set of heuristics shown in Table~\ref{tab:heuristics}. 
For scaling down operations, the scheduler evaluates reducing CPU allocations of single tiers, and batches of tiers, e.g., scaling down 
the $k$ tiers with lowest cpu utilization, $1 < k \leq N$, N being the number of tiers in the microservice graph. 
When scaling up is needed, the scheduler examines the impact of scaling up single tiers, all tiers, or the set of tiers 
that were scaled down in the past $t$ decision intervals, $1 < t < T$ with $T$ chosen empirically. Finally, the scheduler also evaluates the impact of maintaining
the current resource assignment. 

\begin{table}
\caption{\label{tab:heuristics} Resource allocation actions in Sinan. }
%\begin{adjustbox}{width=1.0\linewidth}
\begin{tabular}{cc}
\toprule
\textbf{Category} & \textbf{Actions} \\
\toprule
\hline
\textbf{{Scale Down}}        & Reduce CPU limit of 1 tier \\[0.1cm]
\hdashline[0.5pt/2.5pt]
%\hdashline	
% \rowcolor{lightgray}
	\multirow{2}{*}{\textbf{{Scale Down Batch}}} & Reduce CPU limit of $k$ least utilized tiers, \\[0.1cm]
% \rowcolor{lightgray} 
 & ($1 < k \leq N$) \\
\hdashline[0.5pt/2.5pt]
\textbf{{Hold}} &  Keep current resource allocation \\[0.1cm]
\hdashline[0.5pt/2.5pt]
% \rowcolor{lightgray}
\textbf{Scale Up} & Increase CPU limit of 1 tier \\[0.1cm]
\hdashline[0.5pt/2.5pt]
\textbf{{Scale Up All}}  &  Increase CPU limit of all tiers \\[0.1cm]
\hdashline[0.5pt/2.5pt]
% \rowcolor{lightgray}
\multirow{2}{*}{\textbf{Scale Up Victim}} & Increase CPU limit of recent victim tiers, \\
% \rowcolor{lightgray}
& that are scaled down in previous $t$ cycles \\[0.1cm]
\bottomrule
\end{tabular}
% \vspace{-0.14in}
\end{table}
% \vspace{-0.1in}

The scheduler first excludes operations whose predicted tail latency is higher than $QoS - RMSE_{valid}$. 
Then it uses the predicted violation probability to filter out risky operations, 
with two user-defined thresholds, $p_d$ and $p_u$ ($p_d < p_u$).
These thresholds are similar to those used in autoscaling, where the lower threshold triggers scaling down and 
the higher threshold scaling up; the region between the two thresholds denotes stable operation, where 
the current resource assignment is kept. Specifically, when the violation probability of holding the current assignment 
is smaller than $p_u$, the operation is considered acceptable. Similarly, if there exists a scale down operation 
with violation probability lower than $p_d$, the scale down operation is also considered acceptable. 
When the violation probability of the hold operation is larger than $p_u$, only scaling up operations 
with violation probabilities smaller than $p_u$ are acceptable; if no such actions exist, all tiers are scaled up to their max amount. 
% and as a last resort, when no such scale up operation exists, each tier is scaled up bu maximum amount.
We set $p_u$ such that the validation study's false negatives are no greater than 1\% to eliminate QoS violations, 
and $p_d$ to a value smaller than $p_u$ that favors stable resource allocations, so that resources do not fluctuate 
too frequently unless there are significant fluctuations in utilization and/or user demand. 
Among all acceptable operations, the scheduler selects the one requiring the least resources. % amount. 
%If multiple operations result in the same resource consumption, the scheduler chooses the one with the lowest predicted latency.

The scheduler also has a safety mechanism for cases where the ML model's predicted latency or QoS violation probability deviate 
significantly from the ground truth. %, and where the ML model fails to predict occurring QoS violations. 
If a mispredicted QoS violation occurs, Sinan immediately upscales the resources of all tiers. Additionally, 
given a trust threshold for the model, whenever the number of latency prediction errors or missed QoS violations exceeds the thresholds, 
the scheduler reduces its trust in the model, and becomes more conservative when reclaiming resources. In practice, 
Sinan never had to lower its trust to the ML model. 
%whenever the scheduler spots that under a certain load level, the QoS violation frequency surpasses the threshold, the scheduler disables the model and keeps adding resources to all tiers until no more QoS violation is spotted, and records the final resource allocation. Later, the scheduler will always use the recorded resource configuration for the load until the machine learning models are retrained with the miss-predicted data or even more training samples.

\section{Evaluation}
\label{sec:Evaluation}
% \vspace{-0.05in}

We first evaluate Sinan's accuracy, and training and inference time,
and compare it to other ML approaches. Second, we deploy Sinan %as the resource manager 
on our local cluster, and compare it against autoscaling~\cite{aws_step_scaling}, a widely-deployed empirical technique to manage %the industry standard for managing 
resources in production clouds, and PowerChief~\cite{powerchief}, a resource manager for multi-stage applications that uses queueing analysis.
{\color{black}Third, we show the incremental retraining overheads of Sinan.}
Fourth, we evaluate Sinan's scalability on a large-scale Google Compute Engine (GCE) cluster. 
Finally, we discuss how interpretable ML can improve the management of cloud systems.
% \vspace{-0.08in}

\subsection{Methodology}
%In this section we describe our experiment setup of both local and public cloud experiments.

\noindent \textbf{Benchmarks: } We use the {Hotel Reservation} and {Social Network} benchmarks 
described in Section~\ref{sec:applications}. QoS targets are set with respect to 99\% end-to-end latency, 
200ms for \texttt{Hotel Reservation}, and 500ms for \texttt{Social Network}. 

%\smallskip
\noindent \textbf{Deployment: } Services are deployed with Docker Swarm, with one microservices per container for deployment ease. % of deployment and scaling.
\textit{Locust}~\cite{locust} is used as the workload generator for all experiments. 

%\smallskip
\noindent \textbf{Local cluster: } The cluster has four 80-core servers, with 256GB of RAM each. 
We collected %used our data collection process to aggregate 
31302 and 58499 samples for {Hotel Reservation} and {Social Network} respectively, using our data collection process, 
and split them into training and validation sets with a 9:1 ratio, after random shuffling. {\color{black}The data collection agent runs for 16 hours and 8.7 hours for Social Network and Hotel Reservation respectively, and collecting more training samples do not further improve accuracy.}

\noindent \textbf{GCE cluster: } We use 93 container instances on Google Compute Engine (GCE) to run Social Network, with several replicas per microservice tier. 5900 extra training samples are collected on GCE for the transfer learning.

%{\color{red}- add specs of cluster. }

% \vspace{-0.05in}
\subsection{Sinan's Accuracy and Speed}
%We now evaluate the accuracy and speed of Sinan's ML models on local cluster.

\begin{table*}[]
\caption{RMSE, model size, and performance for three NNs --- {Batch size is 2048. Initial learning rates for MLP, LSTM, and CNN are 0.0001, 0.0005, and 0.001, respectively. All models are trained with a single NVidia Titan Xp}. }
\label{tbl:compare_models}
\centering
%\begin{adjustbox}{width=1.0\linewidth}
\begin{tabular}{@{}ccccccc@{}}
\toprule
\textbf{\large{Apps}} & \textbf{\large{Models}} & \multicolumn{2}{c}{\begin{tabular}[c]{@{}c@{}}\textbf{\large{Train \&Val.}}\\ \textbf{\large{RMSE (ms)}}\end{tabular}} & \begin{tabular}[c]{@{}c@{}}\textbf{\large{Model}} \\ \textbf{\large{size (KB)}}\end{tabular} & \multicolumn{2}{c}{\begin{tabular}[c]{@{}c@{}}\textbf{\large{Train \& Inference}} \\ \textbf{\large{speed (ms/batch)}}\end{tabular}} \\ \toprule
\multirow{3}{*}{\begin{tabular}[c]{@{}c@{}}\textbf{\large{Hotel}} \\ \textbf{\large{Reservation}}\end{tabular}} & \large{MLP} & \textcolor{BrickRed}{\textbf{\large{17.8}}} & \textcolor{BrickRed}{\textbf{\large{18.9}}} & \textcolor{BrickRed}{\textbf{\large{1433}}} & \large{1.9} & \textcolor{BrickRed}{\textbf{\large{3.7}}} \\[0.1cm]
 & \large{LSTM} & \large{17.7} & \large{18.1} & \large{384} & \textcolor{ForestGreen}{\textbf{\large{1.3}}} & \textcolor{ForestGreen}{\textbf{\large{3.2}}} \\[0.1cm]
 & \textbf{\large{CNN}} & \textcolor{ForestGreen}{\textbf{\large{14.2}}} & \textcolor{ForestGreen}{\textbf{\large{14.7}}} & \textcolor{ForestGreen}{\textbf{\large{68}}} & \textcolor{BrickRed}{\textbf{\large{4.5}}} & \large{3.5} \\ \midrule
\multirow{3}{*}{\begin{tabular}[c]{@{}c@{}}\textbf{\large{Social}} \\ \textbf{\large{Network}}\end{tabular}} & \large{MLP} & \textcolor{BrickRed}{\textbf{\large{32.3}}} & \textcolor{BrickRed}{\textbf{\large{34.4}}} & \textcolor{BrickRed}{\textbf{\large{4300}}} & \large{6.4} & \textcolor{BrickRed}{\textbf{\large{5.9}}} \\[0.1cm]
 & \large{LSTM} & \large{29.3} & \large{30.7} & \large{404} & \textcolor{ForestGreen}{\textbf{\large{4.5}}} & \textcolor{ForestGreen}{\textbf{\large{5.6}}}\\[0.1cm]
 & \textbf{\large{CNN}} & \textcolor{ForestGreen}{\textbf{\large{25.9}}} & \textcolor{ForestGreen}{\textbf{\large{26.4}}} & \textcolor{ForestGreen}{\textbf{\large{144}}} & \textcolor{BrickRed}{\textbf{\large{16.0}}} & \large{5.7} \\ \bottomrule
\end{tabular}
%\end{adjustbox}
% \vspace{-0.14in}
\end{table*}

% \begin{figure*}
% \centering
% \begin{minipage}{.88\linewidth}
%   \centering
%   \hspace*{-0.6cm}\includegraphics[width=0.94\linewidth]{figures/violin.pdf}\\
% \end{minipage}%
% \begin{minipage}{.122\linewidth}
%   \centering
%   \hspace*{-0.3cm}\includegraphics[width=0.94\linewidth]{figures/violin_diurnal.pdf}\\
% \end{minipage}\caption{End-to-end latency for \texttt{Social Network} with the three schedulers as load increases. }
% \label{fig:violin}
% \vspace{-0.18in}
% \end{figure*}
% {For now we use the small dedicated cluster to avoid interference from external applications. } % beyond our control. }
Table~\ref{tbl:compare_models} compares the short-term ML model in Sinan (CNN) 
against a multilayer perceptron (MLP), and a long short-term memory (LSTM) network, 
which is traditionally geared towards timeseries predictions. 
We rearrange the system history $X_{RH}$ to be a 2D tensor with shape $T\times(F*N)$, and a 1D vector 
with shape $T*F*N$ for the LSTM and MLP models, respectively. To configure each network's parameters, 
we increase the number of fully-connected, LSTM, and convolutional layers, as well as the number of channels 
in each layer for the MLP, LSTM, and Sinan (CNN), until accuracy levels off. 
Sinan's CNN achieves the lowest RMSE, 
with the smallest model size. Although the CNN is slightly slower than the LSTM, 
its inference latency is within 1\% of the decision interval (1s), which does not delay %impose any delays 
online decisions. %meets the latency requirement of online scheduling. 

Table~\ref{tbl:boosted_tree} shows a similar validation study for the Boosted Trees model. 
Specifically, we quantify the accuracy of anticipating a QoS violation over the next 5 intervals (5s), 
and the number of trees needed for each application. 
% To prevent overfitting, given that the number of data samples 
% with no QoS violation is larger than the ones where QoS is not met by several orders of magnitude, 
% we upsample the second dataset to have a more balanced number of samples. 
For both applications, 
the validation accuracy is higher than 94\%, demonstrating BT's effectiveness in predicting the performance evolution in the near future. 
Sinan always runs on a single NVidia Titan XP GPU with average utilization below 2\%. 
%It is also worth noting that the total training time for 300k samples using boosted trees is within 200s, which is an order of magnitude faster than using neural networks. {\color{red} you need to be careful with this. If boosted trees are that much faster why don't we use them for everything? }

% \vspace{-0.05in}
\subsection{Performance and Resource Efficiency}
\label{sec:online_deployment}

%The services are each deployed on our local cluster described previously. 
%The ML models are again hosted on a GPU NVidia Titan Xp server. 

\begin{table*}[]
\caption{The accuracy, number of trees, and total training time of Boosted Trees using a single NVidia Titan Xp.}
\centering
\label{tbl:boosted_tree}
%\begin{adjustbox}{width=1.0\linewidth}
\begin{tabular}{@{}ccccccc@{}}
\toprule
\textbf{\large{Apps}} & \multicolumn{2}{c}{\begin{tabular}[c]{@{}l@{}}\textbf{\large{Train \& Val.}} \\ \textbf{\large{accuracy (\%)}}\end{tabular}} &
\multicolumn{2}{c}{\begin{tabular}[c]{@{}l@{}}\textbf{\large{Val. false}} \\ \textbf{\large{pos. \& neg.}}\end{tabular}} &
%\begin{tabular}[c]{@{}c@{}}\textbf{\large{Val. false}}\\ \textbf{\large{positive (\%)}}\end{tabular} & \begin{tabular}[c]{@{}c@{}}\textbf{\large{Val. false}} \\ \textbf{\large{negative (\%)}}\end{tabular} & 
\begin{tabular}[c]{@{}c@{}}\textbf{\large{\# of}} \\ \textbf{\large{trees}}\end{tabular} & \begin{tabular}[c]{@{}c@{}}\textbf{\large{Total train}} \\ \textbf{\large{time (s)}}\end{tabular} \\ \toprule
\begin{tabular}[c]{@{}c@{}}\textbf{\large{Hotel}}\\ \textbf{\large{Reservation}}\end{tabular} & \large{94.4} & \large{94.1} & \large{3.2} & \large{3.1} & \large{229} &\large{2.3} \\ \midrule
\begin{tabular}[c]{@{}c@{}}\textbf{\large{Social}}\\ \textbf{\large{Network}}\end{tabular} & \large{95.5}  & \large{94.6} & \large{3.4} & \large{2.0} & \large{239} & \large{6.5} \\ \bottomrule
\end{tabular}
%\end{adjustbox}
% \vspace{-0.14in}
\end{table*}
% \subsubsection{Sensitivity Analysis}

We now evaluate Sinan's ability to reduce resource consumption while meeting QoS on the local cluster. 
We compare Sinan against autoscaling and PowerChief~\cite{powerchief}. We experimented with two autoscaling policies: 
AutoScaleOpt is configured according to~\cite{aws_step_scaling}, which increases resources by 10\% and 30\% when utilization 
is within $[60\%, 70\%)$ and $[70\%, 100\%]$ respectively, and reduces resources by 10\% and 30\% when utilization is within $[30\%, 40\%)$ and $[0\%, 30\%)$. 
AutoScaleCons is more conservative and optimizes for QoS, using thresholds tuned for the examined applications. It increases resources by 10\% and 30\% 
when utilization is within $[30\%, 50\%)$ and $[50\%, 100\%]$, and reduces resources by 10\% when utilization is within $[0\%, 10\%)$. 
PowerChief is implemented as in~\cite{powerchief}, and estimates the queue length and queueing time ahead %in front 
of each tier using network traces obtained through Docker. %'s monitoring framework. 

\begin{figure}[!t]
    \centering
    \begin{tabular}{@{}c@{}}
    \includegraphics[scale=0.35,viewport=50 10 650 160]{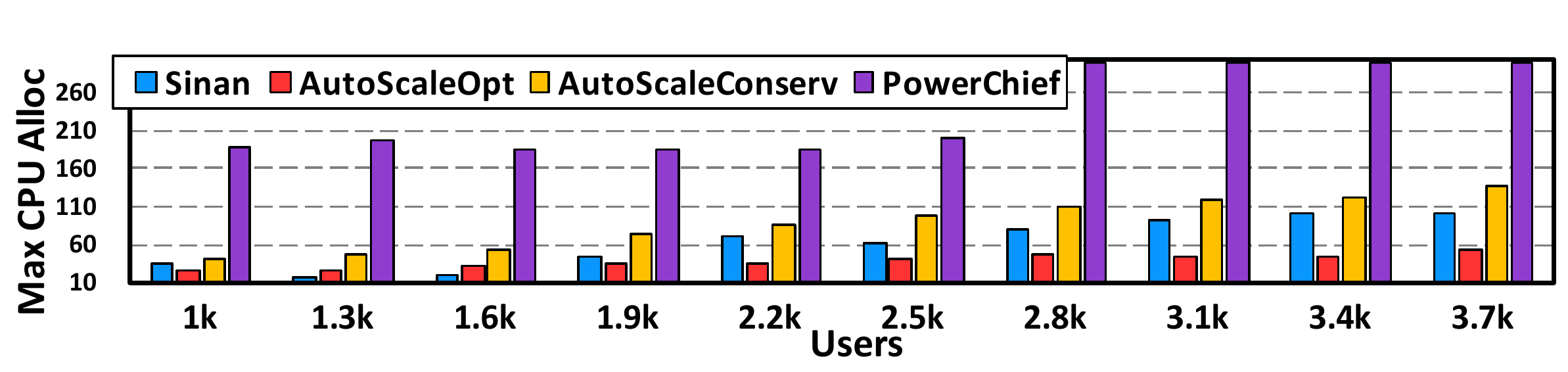}\\
	    %\vspace{-0.32in}
    \includegraphics[scale=0.35, viewport=50 10 650 160]{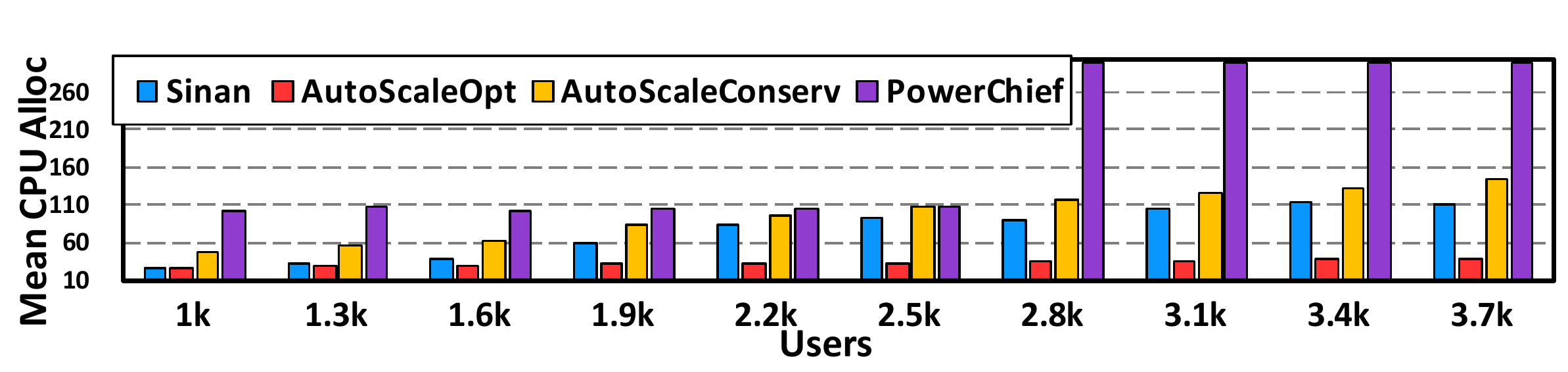}\\
	    %\vspace{-0.06in}
    \includegraphics[scale=0.35, viewport=60 10 650 160]{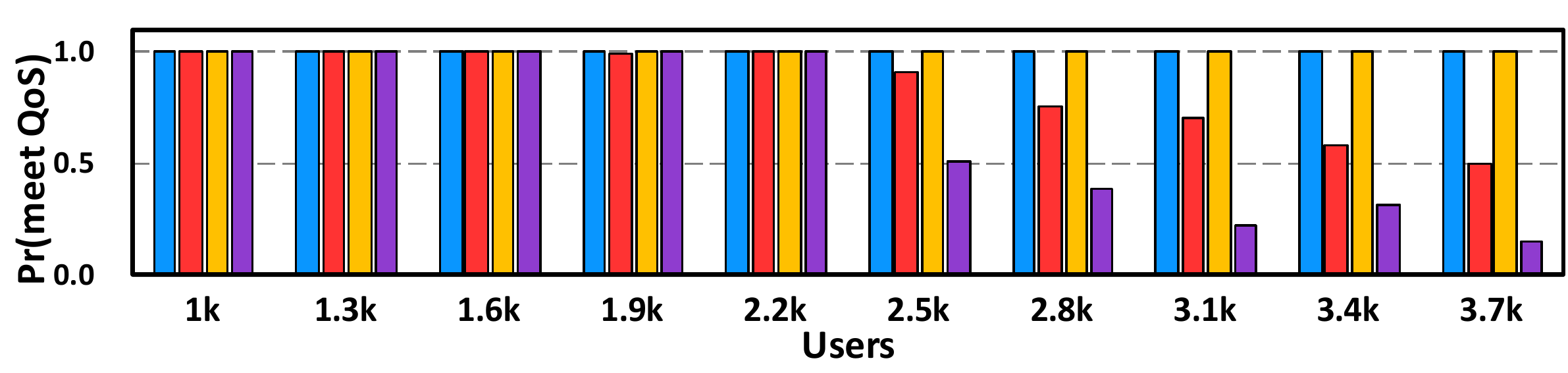}
	   % \vspace{-0.02in}
    \end{tabular}\\
    \small (a) Hotel reservation.
    \begin{tabular}{@{}c@{}}
    \includegraphics[scale=0.35]{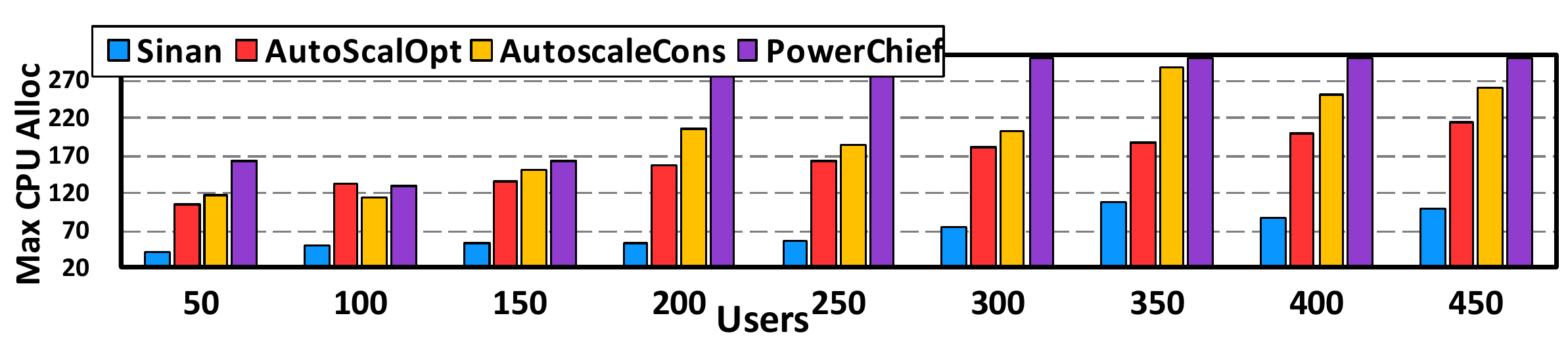}\\
	   % \vspace{-0.06in}
    \includegraphics[scale=0.35]{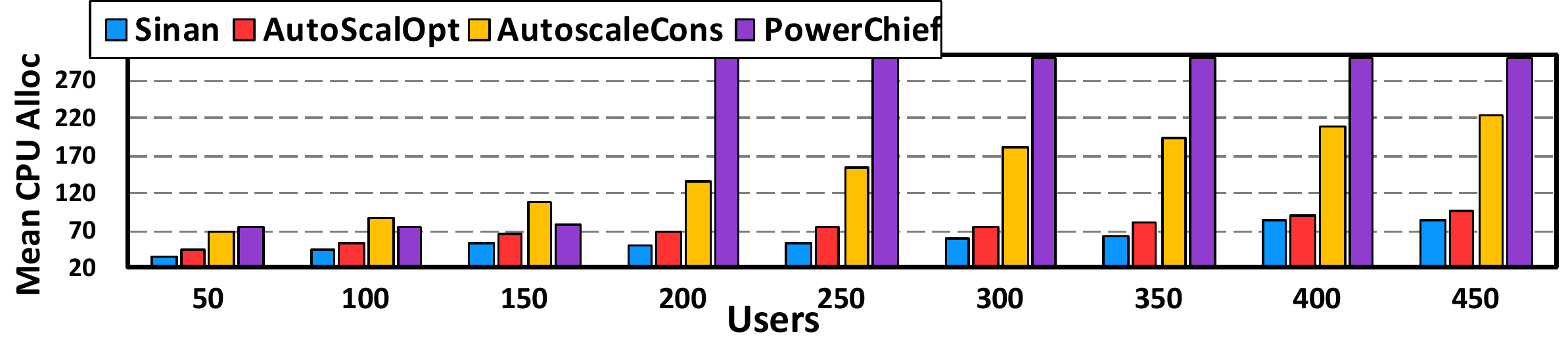}\\
	   % \vspace{-0.06in}
    \includegraphics[scale=0.35]{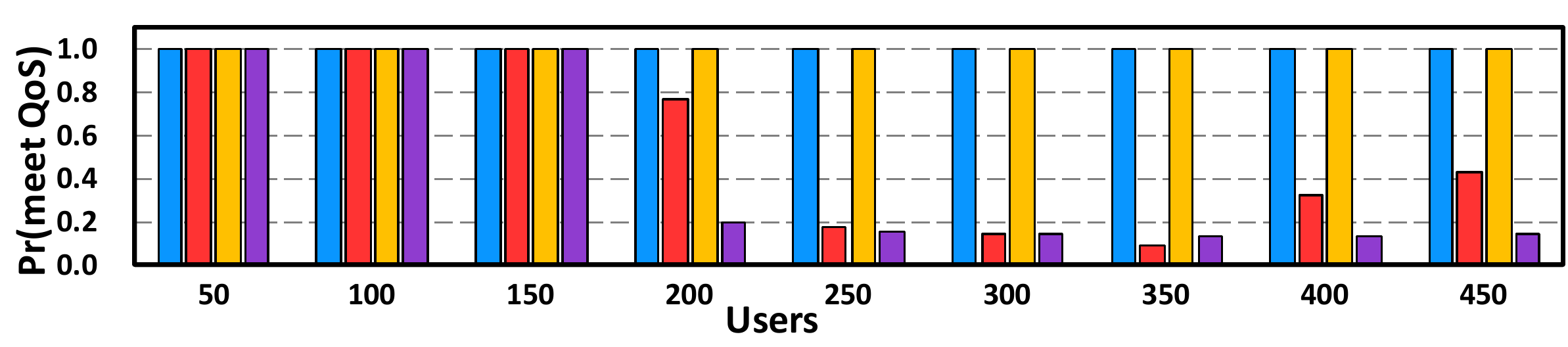}
	   % \vspace{-0.02in}
    \end{tabular}\\
    \small (b) Social network.
    \caption{The mean and max CPU allocation, and the probability of meeting QoS for Sinan, Autoscaling, and PowerChief. }
    \label{fig:load_comparison}
% \vspace{-0.18in}
\end{figure}

For each service, we run 9 experiments with an increasing number of emulated users sending requests under a Poisson distribution with 1 RPS mean arrival rate. 
Figure~\ref{fig:load_comparison} shows the mean and max CPU allocation, and the probability of meeting QoS 
across all studied mechanisms, where CPU allocation is the aggregate number of CPUs assigned to all tiers averaged over time, 
the max CPU allocation is the max of the aggregate CPU allocation over time, %excluding initialization, 
and the probability of meeting QoS is the fraction of execution time when end-to-end QoS is met.
%It's worth mentioncpu limit ing that core and frequency reductions are orthogonal, and in most cases Sinan reduces both the number of cores and the frequency compared to auto-scaling.

\begin{figure*}
%\vspace{-0.18in}
    \centering
    %\small (a) Hotel Reservation under constant and (b) diurnal load.\\
    \begin{tabular}{@{}c@{}}
    \includegraphics[scale=0.4,viewport=100 0 1000 120]{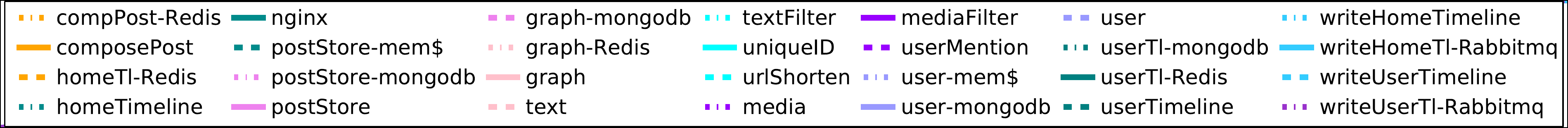}\\
    \includegraphics[scale=0.200,viewport=700 40 1740 700]{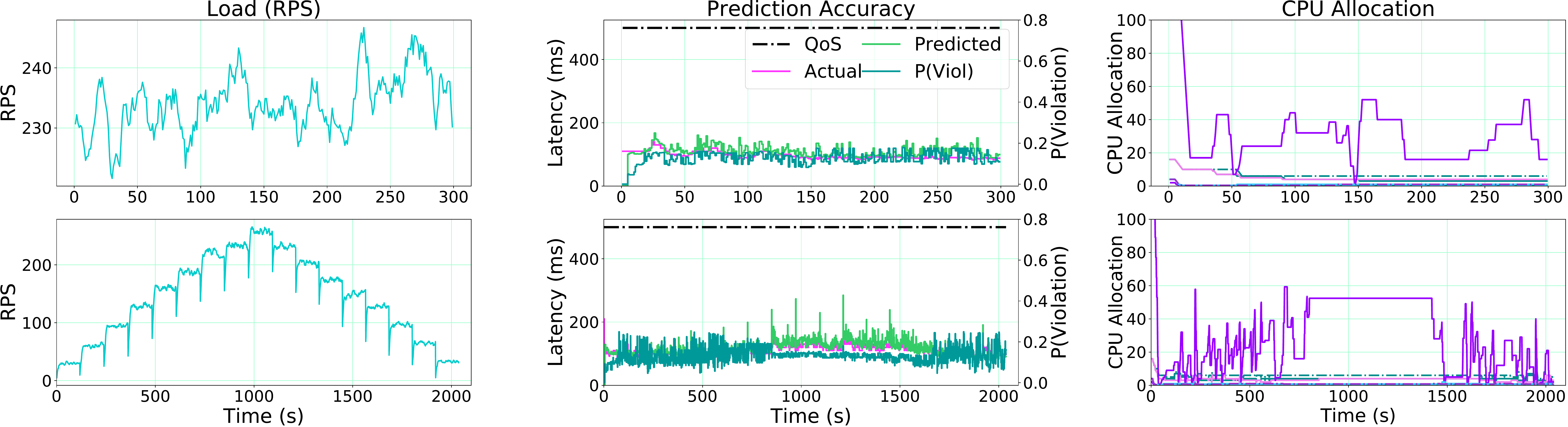}
    \end{tabular}\\
    %\small (c) Social Network under constant and (d) diurnal load.\\
	\caption{\label{fig:detailed_deployment}\textcolor{black}{(Top) RPS, latency, and allocated resources per tier with Sinan for Social Network with 250 users. (Bottom) RPC, latency, and allocated resources per tier with diurnal load. For both scenarios, Sinan's predicted latency closely follows the end-to-end measured latency, avoiding QoS violations and excessive overprovisioning, while allocated resources per tier take into account the impact of microservice dependencies on end-to-end performance. }}
% 	\vspace{-0.08in}
\end{figure*} 
%\end{comment}
For Hotel Reservation, only Sinan and AutoScaleCons meet QoS at all times, 
with Sinan additionally reducing CPU usage by $25.9\%$ on average, and up to $46.0\%$. % compared to AutoScaleCons. 
AutoScaleOpt only meets QoS at low loads, when the number of users is no greater than 1900. 
At 2200 users, AutoScaleOpt starts to violate QoS by 0.7\%, and the probability of 
meeting QoS drops to 90.3\% at 2800 users, and less than 80\% beyond 3000 users. 
Similarly, PowerChief meets QoS for fewer than 2500 users, however the probability of meeting QoS 
drops to 50.8\% at 2800 users, and never exceeds 40\% beyond 3000 users. 
%In terms of resource usage, 
AutoScaleOpt uses 53\% the amount of resources Sinan requires on average, 
at the price of performance unpredictability, %significantly unpredictable performance, 
and PowerChief uses $2.57\times$ more resources than Sinan despite violating QoS. % violations. 
%also introducing considerable QoS violations. 

For the more complicated Social Network, Sinan's performance benefits are more pronounced. 
%Similar to Hotel Reservation, 
Once again, only Sinan and AutoScaleCons meet QoS across loads, while Sinan 
also reduces CPU usage on average by $59.0\%$ and up to $68.1\%$. Both AutoScaleOpt and PowerChief only meet QoS 
for fewer than 150 users, despite using on average $1.26\times$ and up to $3.75\times$ the resources Sinan needs. 
For higher loads, PowerChief's QoS meeting probability is at most 20\% above 150 users, %concurrent users, 
and AutoscaleOpt's QoS meeting probability starts at 76.3\% for 200 users, and decreases to 8.7\% for 350 users. 

%In general, Sinan significantly outperforms previous methods in terms of both reducing resource assignment, and meeting end-to-end QoS.
By reducing both the average and max CPU allocation, Sinan can yield more resources to colocated tasks, 
improving the machine's effective utilization~\cite{Lo14, Delimitrou14, Lo15, chen2019parties}. 
There are three reasons why PowerChief cannot reduce resources similarly and leads to QoS violations. 
First, as discussed in Sec.~\ref{sec:challenges}, the complex topology of microservices means 
that the tier with the longest igress queue, which PowerChief signals as the source of performance issues, is not necessarily the culprit but a symptom. 
Second, in interactive applications, queueing takes place across the system stack, including the NIC, 
OS kernel, network processing, and application, making precise queueing time estimations challenging, especially when tracing uses sampling. 
Finally, the stricter latency targets of microservices, compared to traditional cloud services, 
indicate that small fluctuations in queueing time can result in major QoS violations 
due to imperfect pipelining across tiers causing backpressure to amplify across the system. 
%tracing needs to be entirely transparent to the application, 
%often requiring sampling, which can introduce errors in queue length estimations. 

Fig.~\ref{fig:detailed_deployment} shows the detailed results for Social Network, for 300 concurrent users under a diurnal load. 
The three columns each show requests per second (RPS), predicted latency vs. real latency and predicted QoS violation probability, 
and the realtime CPU allocation. As shown, Sinan's tail latency prediction closely follows the ground truth, 
and is able to react rapidly to fluctuations in the input load.

% \begin{table}[]
% \caption{{Sinan's accuracy on GCE cluster. }}
% % \vspace{-0.22in}
% \label{tbl:scalability}
% \begin{adjustbox}{width=1.0\linewidth}
% \begin{tabular}{@{}cccccccc@{}}
% \toprule
% \textbf{\large{App}} & \textbf{\large{Model}} & \multicolumn{2}{c}{\begin{tabular}[c]{@{}c@{}}\textbf{\large{Train \&Val.}}\\ \textbf{\large{RMSE (ms)}}\end{tabular}} & & \begin{tabular}[c]{@{}c@{}}{\hspace{-0.05in}\textbf{\large{Model size}}} \\ \hspace{-0.05in}\textbf{\large{(KB)}}\end{tabular} & \multicolumn{2}{c}{\begin{tabular}[c]{@{}c@{}}\textbf{\large{Train \& Inference}} \\ \textbf{\large{speed (ms/batch)}}\end{tabular}}\\ \toprule
% \textbf{\large{Social Net}} & \textbf{\large{CNN}} & {{\large{0.93}}} & {{\large{1.47}}} & \multicolumn{2}{c}{{\large{337}}} & {{\large{4500}}} & \large{6.6} \\ 

% %\textbf{\large{App}} & \textbf{\large{Model}} & \multicolumn{2}{c}{\begin{tabular}[c]{@{}l@{}}\textbf{\large{Train \& Val.}} \\ \textbf{\large{accuracy (\%)}}\end{tabular}} &
% %\multicolumn{2}{c}{\begin{tabular}[c]{@{}l@{}}\textbf{\large{Val. false}} \\ \textbf{\large{pos. \& neg.}}\end{tabular}} &
% %\begin{tabular}[c]{@{}c@{}}\textbf{\large{\# of}} \\ \textbf{\large{trees}}\end{tabular} & \begin{tabular}[c]{@{}c@{}}\textbf{\large{Train}} \\ \textbf{\large{time (s)}}\end{tabular} \\ \toprule
% %\textbf{\large{Social Net}} & \textbf{\large{BT}} & \large{96.8}  & \large{98.2} & \large{3.0} & \large{0} & \large{460} & \large{2046} \\ \bottomrule
% \end{tabular}
% \end{adjustbox}
% %\vspace{-0.18in}
% \end{table}

\begin{figure}[!h]
%\vspace{-0.18in}
    \centering
	    \vspace{-0.1in}
    %\small (a) Hotel Reservation under constant and (b) diurnal load.\\
    \begin{tabular}{@{}c@{}}
    \includegraphics[scale=.32]{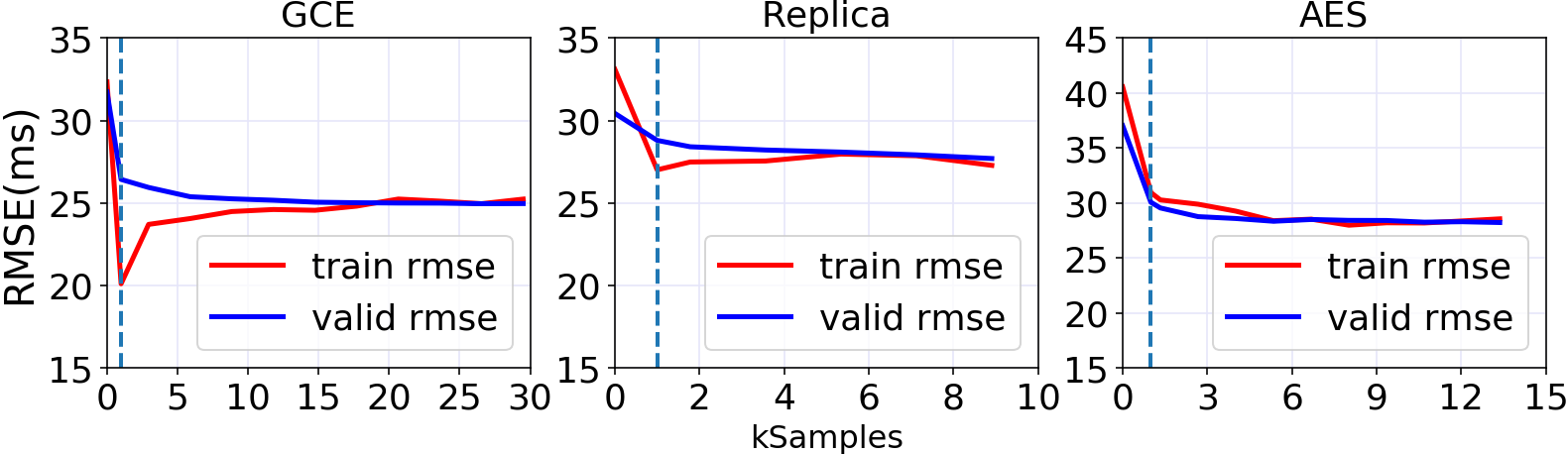}
    \end{tabular}\\
% \vspace{-0.08in}
    %\small (c) Social Network under constant and (d) diurnal load.\\
	\caption{\label{fig:finetune} Training \& validation RMSE of Fine-tunned CNNs with different amounts of samples. }
% \vspace{-0.08in}
\end{figure}

% \vspace{-0.08in}
{\color{black}
\subsection{Incremental Retraining}
\label{sec:finetune}
We show the incremental retraining overheads of Sinan's ML models in three different deployment scenarios with the Social Network applications: 
switching to new server platforms (from the local cluster to a GCE cluster), changing the number of replicas (scale out factor) 
for all microservices except the backend databases (to avoid data migration overheads), 
and modifying the application design by introducing encryption in post messages uploaded by users (posts are encrypted with AES~\cite{daemen1999aes} before being stored in the databases). 
Instead of retraining the ML models from scratch, we use the previously-trained models on the local cluster, 
and fine-tune them using a small amount of newly-collected data, with the initial learning rate $\lambda$ being $1 \times 10^{-5}$, $\frac{1}{100}$ 
of the original $\lambda$ value, in order to preserve the learnt weights in the original model and constrain the new solution derived by the SGD algorithm 
to be in a nearby region of the original one. The results are shown in Fig.~\ref{fig:finetune}, in which the y-axis is the RMSE and the x-axis is 
the number of newly-collected training samples (unit being 1000). The RMSE values with zero new training samples correspond to the original model's 
accuracy on the newly collected training and validation set. In all three scenarios the training and validation RMSE converge, showing that incremental retraining 
in Sinan achieves high accuracy, without the overhead of retraining the entire model from scratch. 
}

{\color{black}
In terms of new server platforms and different replica numbers, the original model already achieve a RMSE of 33.23ms and 33.1ms correspondingly, showing the generalizability of selected input features. The RMSE of original model, when directly applied to the modified application, is higher compared to the two other cases, reaching 40.56ms. In all of the three cases, the validation RMSE is siginificantly reduced with 1000 newly collected training samples (shown by the dotted lines in each figure), which translates to 16.7 minutes of profiling time. The case of GCE, different replica number and modified application stabilize with 5900 samples (1.6 hours of profiling), 1800 samples (0.5 hour of profiling) and 5300 samples (1.5 hours of profiling), and achieve training vs. validation RMSE of 24.8ms vs. 25.2ms, 27.5ms vs. 28.2ms, and 28.4ms vs. 28.3ms correspondingly.
}

% \vspace{-0.08in}
\subsection{Sinan's Scalability}

We now show Sinan's scalability on GCE running Social Network. We use the fine-tuned model described in Section~\ref{sec:finetune}. Apart from the CNN, XGBoost achieves training and validation accuracy of 96.1\% and 95.0\%. The model's size and speed 
remain unchanged, since they share the same architecture with the local cluster models. 

\begin{figure}[!t]
    \begin{tabular}{@{}c@{}}
	    \includegraphics[scale=0.364,viewport=10 10 800 140]{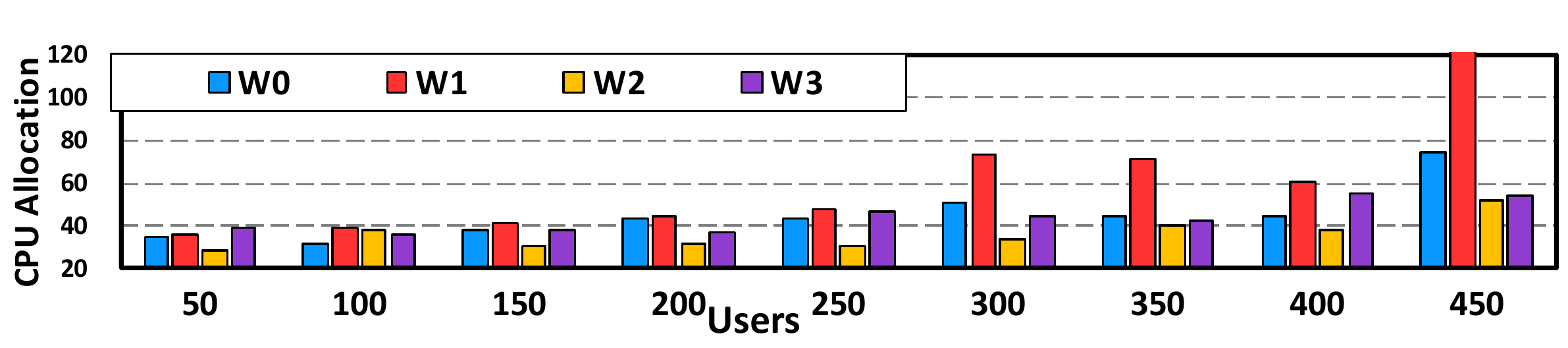}
    \end{tabular}\\
    % \vspace{-0.06in}
    \caption{Comparison of the average CPU allocation of four request mixes for Social Network on GCE. }
    \label{fig:gcp_cpu}
% \vspace{-0.1in}
\end{figure}

\begin{figure}[!h]
    \centering
	   % \vspace{-0.16in}
    %\small (a) Hotel Reservation under constant and (b) diurnal load.\\
    \begin{tabular}{@{}c@{}}
    \includegraphics[scale=0.35]{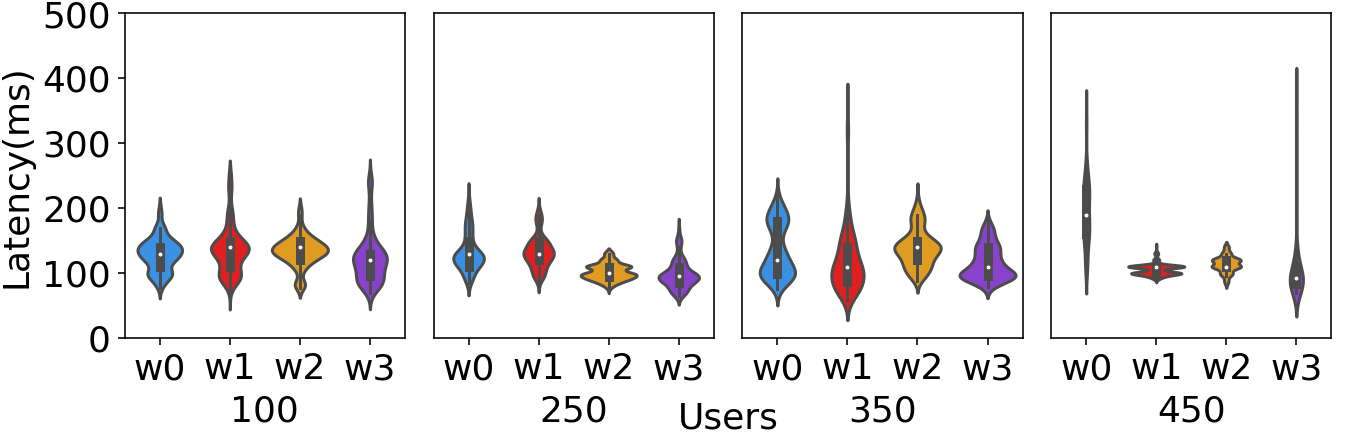}
    \end{tabular}\\
    %\small (c) Social Network under constant and (d) diurnal load.\\
	\caption{\label{fig:gce_lat_violin} 99$^{th}$ percentile latency distribution for four workload types of Social Network on GCE, managed by Sinan. }
\end{figure}

To further test Sinan's robustness to workload changes, we experimented with four %different 
workloads for Social Network, by varying request types. Some requests, like ComposePost involve the majority of microservices, and hence are more resource intensive, 
while others, like ReadUserTimeline involve a much smaller number of tiers, and are easier to allocate resources for. 
We vary the ratio of ComposePost:ReadHomeTimeline:ReadUserTimeline requests; the ratios of the $W0$, $W1$, $W2$ and $W3$ workloads 
are 5:80:15, 10:80:10, 1:90:9, and 5:70:25, where $W0$ has the same ratio as the training set. The ratios are representative of 
different social media engagement scenarios~\cite{nr}. The average CPU allocation and tail latency distribution are shown 
in Fig.~\ref{fig:gcp_cpu} and Fig.~\ref{fig:gce_lat_violin}. Sinan always meets QoS, adjusting resources accordingly. 
$W1$ requires the max compute resources (170 vCPUs for 450 users), because of the highest number of ComposePost requests, 
which trigger compute-intensive ML microservices. 

% \vspace{-0.08in}
\subsection{Explainable ML}

%\begin{comment}
%\begin{wrapfigure}[40]{r}[\dimexpr\columnwidth+\columnsep\relax]{13.5cm}
% \begin{wrapfigure}[38]{r}[\dimexpr\columnwidth+\columnsep\relax]{12.3cm}
% 	\vspace{-0.18in}

For users to trust ML, it is important to interpret its output with respect 
to the system it manages, instead of treating ML as a black box. We are specifically 
interested in understanding what makes some features in the model %prediction process being
 more important than others. The benefits %of understanding this %the importance of input features 
 are threefold: 1) debugging the models; 2) identifying and fixing performance issues; 
 3) filtering out spurious features to reduce model size and speed up inference.

	\subsubsection{Interpretability methods}\vspace{0.08in}%\hfill\hspace{4cm}\hfill

For the CNN model, we adopt the widely-used ML interpretability approach LIME~\cite{lime}. 
LIME interprets NNs by identifying their key input features which contribute most to predictions.
Given an input $X$, LIME perturbs $X$ to obtain a set of artificial samples which are close to $X$ in the feature space.
%For boosted trees, at each split point, the boosting algorithm chooses one of the input features which has the largest entropy gain among all features as the splitting criterion. As a result, the times that each feature is used as the splitting criterion indicates the importance of the feature. We can leverage this information to figure out the importance of the resources for each tier.
Then, LIME classifies the perturbed samples with the NN, and uses the labeled data 
to fit a linear regression model. Given that linear regression is easy to interpret, LIME uses it 
to identify important features based on the regression parameters. Since we are mainly interested 
in understanding the culprit of the QoS violations, we choose samples $X$ from the timesteps 
where QoS violations occur. We perturb the features of a given tier or resource by multiplying that feature 
with different constants. For example, to study the importance of MongoDB, 
we multiply its utilization history with two constants 0.5 and 0.7, and generate multiple perturbed samples. 
Then, we construct a dataset with all perturbed and original data to train the linear regression model. 
Last, we rank the importance of each feature by summing the value of their associated weights.

	\subsubsection{Interpreting the CNN}%\hspace{5cm}\hfill
\begin{figure}[!h]
%\vspace{-0.18in}
    \centering
	   % \vspace{-0.16in}
    %\small (a) Hotel Reservation under constant and (b) diurnal load.\\
    \begin{tabular}{@{}c@{}}
    \includegraphics[scale=0.18,viewport=270 45 600 625]{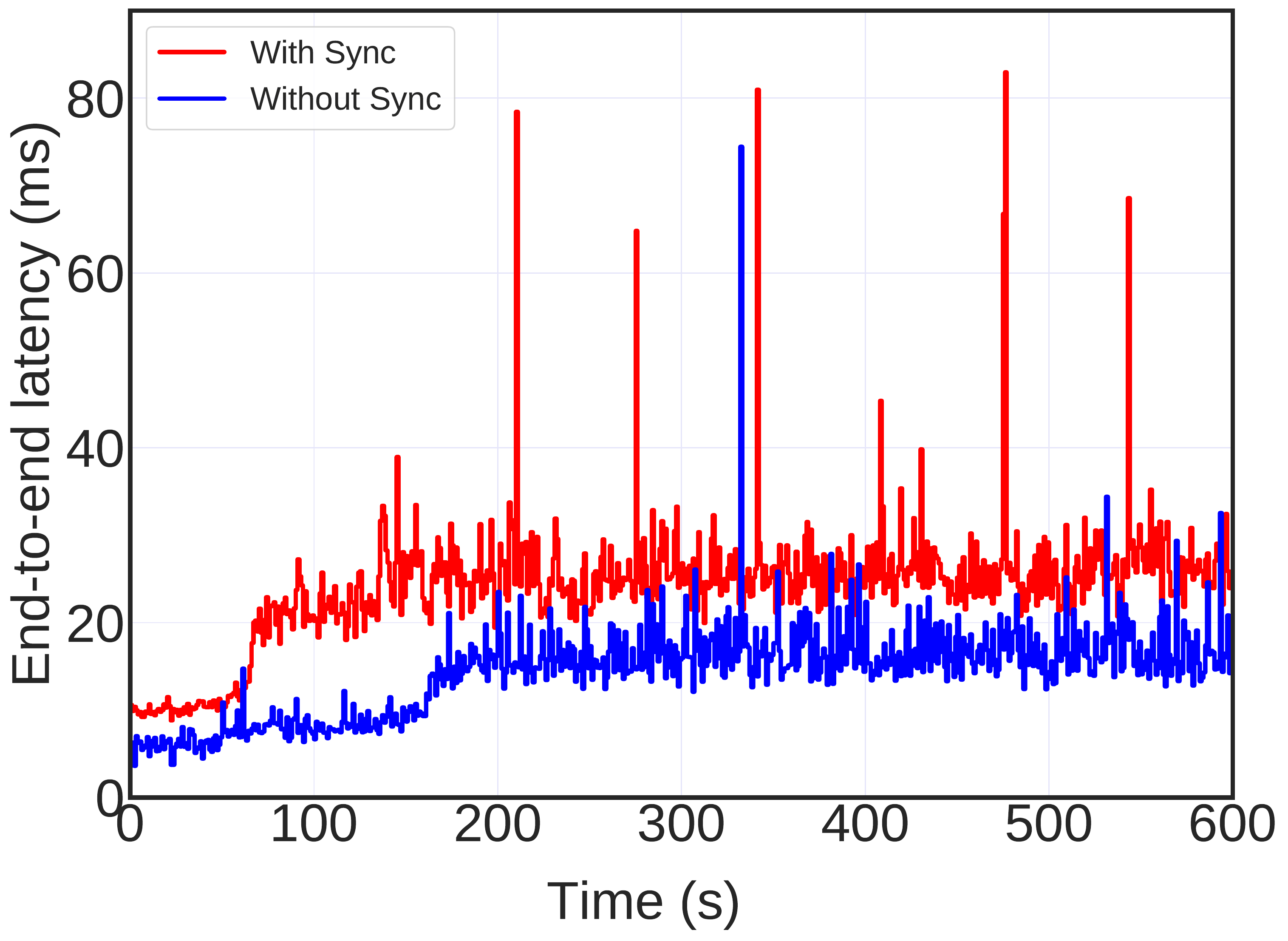}
    \end{tabular}\\
    % \vspace{-0.08in}
	\caption{\label{fig:compare} \textcolor{black}{Tail latency for the Social Network application when Redis's logging is enabled (red) and disabled (black). Sinan identified Redis as the source of unpredictable performance, and additionally determined the resources that were being saturated, pointing to the issue being in Redis's logging functionality. Disabling logging significanly improved performance, which is also reflected in that tier's importance, as far as meeting QoS is concerned, being reduced. }}
% \vspace{-0.08in}
\end{figure}

We used LIME to correct performance issues in Social Network~\cite{gan2019open}, 
where tail latency experienced periods of spikes and instability despite the low load, 
as shown by the red line in Fig.~\ref{fig:compare}. Manual debugging is cumbersome, as it requires 
delving into each tier, and potentially combinations of tiers to identify the root cause. Instead, 
we leverage explainable ML to filter the search space. First, we identify the top-5 most important tiers; 
the results are shown in the w/ Sync part of Table~\ref{tab:exp_ml}. We find that 
the most important tier for the model's prediction is social-graph Redis, instead of tiers with heavy CPU utilization, like \texttt{nginx}.

\begin{table*}[]
\caption{Top-5 most critical tiers and resources for QoS with/without log synchronization in Social Network --- SGrf and WUsr are social graph and write user, respectively.}
\label{tab:exp_ml}
\centering
%\begin{adjustbox}{width=1.0\linewidth}
\begin{tabular}{@{}cccccccc@{}}
\toprule
\multirow{4}{*}{\begin{tabular}[c]{@{}c@{}}\\\\\textbf{\large{w/}}\\ \textbf{\large{Sync}}\end{tabular}} & \textbf{\large{Tiers}} & \begin{tabular}[c]{@{}c@{}}\textbf{\large{\textcolor{BrickRed}{SGrf}}}\\ \textbf{\large\textcolor{BrickRed}{{Redis}}}\end{tabular} & \begin{tabular}[c]{@{}c@{}}\textbf{\large{post}}\\ \textbf{\large{storage}}\end{tabular}  &  \begin{tabular}[c]{@{}c@{}}\textbf{\large{WUsr}}\\ \textbf{\large{timeline}}\end{tabular} & \begin{tabular}[c]{@{}c@{}}\textbf{\large{SGrf}}\\ \textbf{\large{MongoDB}}\end{tabular} & \textbf{\large{SGrf}}  \\[0.1cm] \cmidrule(l){2-8} 
 & \textbf{\large{Weights}} & \large{5109.9} & \large{1609.8} & \large{1503.1} & \large{849.7} & \large{482.7} &\\[0.1cm] \cmidrule(l){2-8}
%\multirow{2}{*}{\begin{tabular}[c]{@{}c@{}}\textbf{\large{Top-5}}\\ 
%\textbf{\large{resources}}\end{tabular}} & 
& \begin{tabular}[c]{@{}c@{}}\textbf{\large{Resource}}\\ \textbf{\large{utilization}}\end{tabular} & \begin{tabular}[c]{@{}c@{}}\textbf{\large{cache}}\\ \textbf{\large{memory}}\end{tabular}  & \textbf{\large{RSS}} & \textbf{\large{\# of cores}} & \begin{tabular}[c]{@{}c@{}}\textbf{\large{CPU}}\\ \textbf{\large{utilization}}\end{tabular} & \begin{tabular}[c]{@{}c@{}}\textbf{\large{received}}\\ \textbf{\large{packets}}\end{tabular} \\[0.1cm] \cmidrule(l){2-8} 
 & \textbf{\large{Weights}} & \large{15181.9} & \large{1576.1} & \large{658.5} & \large{322.7} & \large{20.0} &\\[0.1cm]\toprule
\multirow{2}{*}{\begin{tabular}[c]{@{}c@{}}\textbf{\large{w/o}}\\ 
\textbf{\large{Sync}}\end{tabular}} & 
\textbf{\large{Tiers}} & \begin{tabular}[c]{@{}c@{}}\textbf{\large{WUsr}}\\ \textbf{\large{timeline}}\end{tabular} & \begin{tabular}[c]{@{}c@{}}\textbf{\large{WUsr}}\\ \textbf{\large{rabbitmq}}\end{tabular} & \begin{tabular}[c]{@{}c@{}}\textbf{\large{SGrf}}\\ \textbf{\large{MongoDB}}\end{tabular} & \textbf{\large{SGrf}} & \begin{tabular}[c]{@{}c@{}}\textbf{\large\textcolor{BrickRed}{{SGrf}}}\\ \textbf{\large\textcolor{BrickRed}{{Redis}}}\end{tabular} \\[0.1cm] \cmidrule(l){2-8} 
 & \textbf{\large{Weights}} & \large{3948.6} & \large{3601.6} & \large{1794.0} & \large{600.9} & \large{451.7} &\\
 \bottomrule
\end{tabular}
%\end{adjustbox}
% \vspace{-0.08in}
\end{table*}

% \begin{wrapfigure}[10]{r}{0.25\textwidth}
%     \centering
%     \includegraphics[scale=0.14,viewport=270 0 600 560]{figures/compare.pdf}
%   \vspace{-0.08in}
%   \caption{Social Network tail latency with/without DB logging.}
%   \label{fig:compare}
% \end{wrapfigure}

We then examine the importance of each resource metric for Redis, 
and find that the most meaningful resources are cache and resident working set size, 
which correspond to data from disk cached in memory, and non-cached memory, % for a process, 
including stacks and heaps. Using these hints, we check the memory configuration and statistics 
of Redis, and identify that it is set to record logs in persistent storage every minute. 
For each operation, Redis forks a new process and copies all written memory to disk; 
during this it stops serving requests. 
%and Redis stops serving requests during that period.

%\begin{wrapfigure}[12]{r}{0.3\textwidth}
%    \centering 
%    \vspace{-0.5cm}
%  \includegraphics[scale=0.16]{figures/explain.png}
%    \vspace{-0.4cm}
%    \caption{Perturbing the input by multiplying a tier or resource usage with a constant mask.}
%    \label{fig:perturb}
%\end{wrapfigure}

Disabling the log persistence eliminated most of the latency spikes, as shown by the black line in Fig.~\ref{fig:compare}. 
We further analyze feature importance in the model trained with data from the modified Social Network, 
and find that the importance of social-graph Redis is significantly reduced, 
as shown in the w/o Sync part of Table~\ref{tab:exp_ml}, in agreement with our observation 
that the service's tail latency is no longer sensitive to that tier. %'s operation.

\section{Related Work}
\label{sec:RelatedWork}

We now review related work on microservices, cloud management, and the use of machine learning in cloud systems. 

\vspace{0.06in}
\noindent{\bf{Microservices: }} The emergence of microservices has prompted recent
work to study their characteristics and system implications~\cite{gan2019open,Gan18,Rahman19}. %In terms of benchmarks, 
DeathstarBench~\cite{gan2019open} and uSuite~\cite{sriraman2018mu} are two representative 
microservice benchmark suites. \textcolor{black}{DeathStarBench includes several end-to-end applications built with microservices, 
and explores the system implications of microservices in terms of server design, network and OS overheads, 
cluster management, programming frameworks, and tail at scale effects. uSuite also introduces a number of 
multi-tier applications built with microservices and studies their performance and resource characteristics. 
Urgaonkar et al.~\cite{Urgaonkar05} introduced analytical modeling to multi-tier applications, which
accurately captured the impact of aspects like concurrency limits and caching policies. 
The takeaway of all these studies is that, despite their benefits, microservices change several assumptions 
current cloud infrastructures are designed with, introducing new system challenges both in hardware and software. }

% Urgaonkar et al.~\cite{Urgaonkar05} introduced analytical modeling to multi-tier applications. 
%which 
% accurately captured the impact of aspects like concurrency limits and caching policies. 
In terms of resource management, Wechat~\cite{zhou2018overload} manages microservices with overload control, by matching the throughput 
of the upstream and downstream services; PowerChief~\cite{powerchief} dynamically power boosts bottleneck services in multi-phase 
applications, and Suresh et al.~\cite{suresh2017distributed} leverage overload control 
and adopt deadline-based scheduling to improve tail latency in multi-tier workloads. Finally, Sriraman et al.~\cite{Sriraman18} 
present an autotuning framework for microservice concurrency, and show the impact of threading decisions on application performance and responsiveness. 
%while Rahman et al.~\cite{Rahman19} show that analytical models can be used to predict the performance of multi-tier applications with different resource allocations. 
 
\vspace{0.06in}
\noindent{\bf{Cloud resource management: }} 
The prevalence of cloud computing has motivated many cluster management designs. 
Quasar~\cite{Delimitrou14,Delimitrou14b}, Mesos~\cite{Mesos11}, Torque~\cite{Torque}, and Omega~\cite{omega13} 
all target resource allocation in large, multi-tenant clusters. Quasar~\cite{Delimitrou14} is 
a QoS-aware cluster manager that leverages machine learning to identify the resource preferences 
of new, unknown applications, and allocate resources in a way that meets their performance requirements without 
sacrificing resource efficiency. Mesos~\cite{Mesos11} 
is a two-level scheduler that makes resource offers to different tenants, while 
Omega~\cite{omega13} uses a shared-state approach to scale to larger clusters. 
More recently, PARTIES~\cite{chen2019parties} leveraged the intuition that resources 
are fungible to co-locate multiple interactive services on a server, using resource partitioning. 
Autoscaling~\cite{qu2018auto} is the industry standard 
for elastically scaling allocations based on utilization~\cite{Autoscale,AutoscaleLimit,Chase01}. \textcolor{black}{While all 
these systems improve the performance and/or resource efficiency of the cloud infrastructures they manage, 
they are designed for monolithic applications, or services with a few tiers, and cannot be directly applied to microservices. }
%However, this work targets single-tier batch and interactive jobs, and does not account for the impact of microservices dependencies. 

\vspace{0.06in}
\noindent{\bf{ML in cloud systems: }} There has been growing interest in leveraging ML %techniques 
to tackle system problems, especially resource management. Quasar leverages collaborative filtering to identify appropriate resource allocations for unknown jobs. 
Autopilot~\cite{rzadca2020autopilot} uses an ensemble of models to infer efficient CPU and memory job configurations. 
%the efficient cpu and memory configuration of jobs.
Resource central~\cite{cortez2017resource} characterizes VM instance behavior 
and trains a set of ML models offline, which accurately predict CPU utilization, deployment size, lifetime, etc. 
using random forests and boosting trees. %Mao et al.~\cite{Mao18} recently showed that ML can also be beneficial in scheduling for clusters hosting data processing frameworks. 
\textcolor{black}{Finally, Seer~\cite{gan2018seer} presented a performance debugging system for microservices, which leverages deep learning 
to identify patterns of common performance issues and to help locate and resolve them. }

\section{Conclusion}
\label{sec:Conclusions}

We have presented Sinan, a scalable and QoS-aware resource manager for interactive microservices. 
Sinan highlights the challenges of managing complex microservices, and leverages a set of 
validated ML models to infer the impact allocations have on end-to-end tail latency. 
Sinan operates online and adjusts its decisions to account for application changes. 
We have evaluated Sinan both on local clusters and public clouds GCE) 
across different microservices, %over different end-to-end applications built with microservices, 
and showed that it meets QoS without sacrificing resource efficiency. Sinan highlights 
the importance of automated, data-driven approaches that manage the cloud's complexity in a practical way.

%\vspace{0.14in}
%\noindent{\large{\bf{Acknowledgements}}}
%\vspace{0.03in}
%\section*{Acknowledgements}

%We sincerely thank all the participants in the user study for their time and effort.
%We also thank Daniel Sanchez, Ed Suh, Grant Ayers, Mingyu Gao, Ana Klimovic, and the rest of the MAST group,
%as well as the anonymous reviewers for their feedback on earlier versions of this manuscript.
%This work was supported by the Stanford Platform Lab, NSF
%grant CNS-1422088, and a John and Norma Balen Sesquicentennial Faculty Fellowship.
%
% The acknowledgments section is defined using the "acks" environment (and NOT an unnumbered section). This ensures
% the proper identification of the section in the article metadata, and the consistent spelling of the heading.

%\clearpage
\balance
%

% The next two lines define the bibliography style to be used, and the bibliography file.
\bibliographystyle{IEEEtranS}
\bibliography{references}

\begin{thebibliography}{10}

\bibitem{twitter_decomposing}
Decomposing twitter: Adventures in service-oriented architecture.
\newblock
  {\url{https://www.slideshare.net/InfoQ/decomposing-twitter-adventures-in-serviceoriented-architecture}}.

\bibitem{docker}
Docker containers.
\newblock \url{https://www.docker.com/}.

\bibitem{locust}
Locust.
\newblock \url{https://locust.io/}.

\bibitem{aws_step_scaling}
Step and simple scaling policies for amazon ec2 auto scaling.
\newblock
  \url{https://docs.aws.amazon.com/autoscaling/ec2/userguide/as-scaling-simple-step.html}.

\bibitem{grpc}
Why grpc?
\newblock \url{https://grpc.io/}.

\bibitem{Cockroft16}
The evolution of microservices.
\newblock
  \url{https://www.slideshare.net/adriancockcroft/evolution-of-microservices-craft-conference},
  2016.

\bibitem{Cockroft15}
Microservices workshop: Why, what, and how to get there.
\newblock
  \url{http://www.slideshare.net/adriancockcroft/microservices-workshop-craft-conference}.

\bibitem{Autoscale}
Autoscale.
\newblock \url{https://cwiki.apache.org/cloudstack/autoscaling.html}.

\bibitem{AutoscaleLimit}
Aws autoscaling.
\newblock \url{http://aws.amazon.com/autoscaling/}.

\bibitem{Chase01}
Jeffrey Chase, Darrell Anderson, Prachi Thakar, Amin Vahdat, and Ronald Doyle.
\newblock Managing energy and server resources in hosting centers.
\newblock In {\em Proceedings of SOSP}. Banff, CA, 2001.

\bibitem{chen2019parties}
Shuang Chen, Christina Delimitrou, and Jos{\'e}~F Mart{\'\i}nez.
\newblock Parties: Qos-aware resource partitioning for multiple interactive
  services.
\newblock In {\em Proceedings of the Twenty-Fourth International Conference on
  Architectural Support for Programming Languages and Operating Systems}, pages
  107--120. ACM, 2019.

\bibitem{xgboost}
Tianqi Chen and Carlos Guestrin.
\newblock {XGBoost}: A scalable tree boosting system.
\newblock In {\em Proceedings of the 22nd ACM SIGKDD International Conference
  on Knowledge Discovery and Data Mining}, KDD '16, pages 785--794, New York,
  NY, USA, 2016. ACM.

\bibitem{mxnet}
Tianqi Chen, Mu~Li, Yutian Li, Min Lin, Naiyan Wang, Minjie Wang, Tianjun Xiao,
  Bing Xu, Chiyuan Zhang, and Zheng Zhang.
\newblock Mxnet: {A} flexible and efficient machine learning library for
  heterogeneous distributed systems.
\newblock {\em CoRR}, abs/1512.01274, 2015.

\bibitem{cortez2017resource}
Eli Cortez, Anand Bonde, Alexandre Muzio, Mark Russinovich, Marcus Fontoura,
  and Ricardo Bianchini.
\newblock Resource central: Understanding and predicting workloads for improved
  resource management in large cloud platforms.
\newblock In {\em Proceedings of the 26th Symposium on Operating Systems
  Principles}, pages 153--167. ACM, 2017.

\bibitem{daemen1999aes}
Joan Daemen and Vincent Rijmen.
\newblock Aes proposal: Rijndael.
\newblock 1999.

\bibitem{Delimitrou13}
Christina Delimitrou and Christos Kozyrakis.
\newblock {Paragon: QoS-aware scheduling for heterogeneous datacenters}.
\newblock In {\em Proceedings of the Eighteenth International Conference on
  Architectural Support for Programming Languages and Operating Systems
  (ASPLOS)}. 2013. 

\bibitem{Delimitrou13e}
Christina Delimitrou and Christos Kozyrakis.
\newblock {QoS-Aware Admission Control in Heterogeneous Datacenters}.
\newblock In {\em Proceedings of the International Conference of Autonomic Computing (ICAC)}. 2013. 

\bibitem{Delimitrou14}
Christina Delimitrou and Christos Kozyrakis.
\newblock {Quasar: Resource-Efficient and QoS-Aware Cluster Management}.
\newblock In {\em Proceedings of the Nineteenth International Conference on
  Architectural Support for Programming Languages and Operating Systems
  (ASPLOS)}. Salt Lake City, UT, USA, 2014.

\bibitem{Delimitrou15}
Christina Delimitrou and Daniel Sanchez and Christos Kozyrakis.
\newblock {Tarcil: Reconciling Scheduling Speed and Quality in Large Shared Clusters}.
\newblock In {\em Proceedings of the Sixth ACM Symposium on Cloud Computing (SOCC)}. 2015. 

\bibitem{Delimitrou16}
Christina Delimitrou and Christos Kozyrakis.
\newblock {HCloud: Resource-Efficient Provisioning in Shared Cloud Systems}.
\newblock In {\em Proceedings of the Twenty First International Conference on
  Architectural Support for Programming Languages and Operating Systems
  (ASPLOS)}. 2016.

\bibitem{Delimitrou17}
Christina Delimitrou and Christos Kozyrakis.
\newblock {Bolt: I Know What You Did Last Summer... In The Cloud}.
\newblock In {\em Proceedings of the Twenty Second International Conference on
  Architectural Support for Programming Languages and Operating Systems
  (ASPLOS)}. 2017.

\bibitem{Gan18}
Yu Gan, Meghna Pancholi, Dailun Cheng, Siyuan Hu, Yuan He, and Christina
  Delimitrou.
\newblock Seer: leveraging big data to navigate the complexity of cloud
  debugging.
\newblock In {\em HotCloud}, 2018.

\bibitem{Gan18b}
Yu Gan and Christina Delimitrou.
\newblock The architectural implications of cloud microservices.
\newblock In {\em Computer Architecture Letters}, 2018.

\bibitem{Delimitrou14b}
Christina Delimitrou and Christos Kozyrakis.
\newblock {QoS-Aware Scheduling in Heterogeneous Datacenters with Paragon}.
\newblock In {\em ACM Transactions on Computer Systems (TOCS)}. 2014.

\bibitem{denning1968working}
Peter~J Denning.
\newblock The working set model for program behavior.
\newblock {\em Communications of the ACM}, 11(5):323--333, 1968.

\bibitem{gan2018seer}
Yu~Gan, Meghna Pancholi, Dailun Cheng, Siyuan Hu, Yuan He, and Christina
  Delimitrou.
\newblock Seer: leveraging big data to navigate the complexity of cloud
  debugging.
\newblock In {\em Proceedings of the 10th USENIX Conference on Hot Topics in
  Cloud Computing}, pages 13--13. USENIX Association, 2018.

\bibitem{gan2019open}
Yu~Gan, Yanqi Zhang, Dailun Cheng, Ankitha Shetty, Priyal Rathi, Nayan Katarki,
  Ariana Bruno, Justin Hu, Brian Ritchken, Brendon Jackson, Kelvin Hu, Meghna
  Pancholi, Brett Clancy, Chris Colen, Fukang Wen, Catherine Leung, Siyuan
  Wang, Leon Zaruvinsky, Mateo Espinosa, Yuan He, and Christina Delimitrou.
\newblock An open-source benchmark suite for microservices and their
  hardware-software implications for cloud \& edge systems.
\newblock In {\em Proceedings of the Twenty-Fourth International Conference on
  Architectural Support for Programming Languages and Operating Systems}, pages
  3--18. ACM, 2019.

\bibitem{gittins2011multi}
John Gittins, Kevin Glazebrook, and Richard Weber.
\newblock {\em Multi-armed bandit allocation indices}.
\newblock John Wiley \& Sons, 2011.

\bibitem{gligoric2018constraints}
Kristina Gligori{\'c}, Ashton Anderson, and Robert West.
\newblock How constraints affect content: The case of twitter’s switch from
  140 to 280 characters.
\newblock In {\em Twelfth International AAAI Conference on Web and Social
  Media}, 2018.

\bibitem{Mesos11}
Ben Hindman, Andy Konwinski, Matei Zaharia, Ali Ghodsi, Anthony~D. Joseph,
  Randy Katz, Scott Shenker, and Ion Stoica.
\newblock Mesos: A platform for fine-grained resource sharing in the data
  center.
\newblock In {\em Proceedings of NSDI}. Boston, MA, 2011.

\bibitem{kwak2010twitter}
Haewoon Kwak, Changhyun Lee, Hosung Park, and Sue Moon.
\newblock What is twitter, a social network or a news media?
\newblock In {\em Proceedings of the 19th international conference on World
  wide web}, pages 591--600. AcM, 2010.

\bibitem{Lin11}
Ching-Chi Lin, Pangfeng Liu, and Jan-Jan Wu.
\newblock Energy-aware virtual machine dynamic provision and scheduling for
  cloud computing.
\newblock In {\em Proceedings of the 2011 IEEE 4th International Conference on
  Cloud Computing (CLOUD)}. Washington, DC, USA, 2011.

\bibitem{Lo14}
David Lo, Liqun Cheng, Rama Govindaraju, Luiz~Andr{\'e} Barroso, and Christos
  Kozyrakis.
\newblock Towards energy proportionality for large-scale latency-critical
  workloads.
\newblock In {\em Proceedings of the 41st Annual International Symposium on
  Computer Architecuture (ISCA)}. Minneapolis, MN, 2014.

\bibitem{Lo15}
David Lo, Liqun Cheng, Rama Govindaraju, Parthasarathy Ranganathan, and
  Christos Kozyrakis.
\newblock Heracles: Improving resource efficiency at scale.
\newblock In {\em Proc. of the 42Nd Annual International Symposium on Computer
  Architecture (ISCA)}. Portland, OR, 2015.

\bibitem{boosting}
Llew Mason, Jonathan Baxter, Peter Bartlett, and Marcus Frean.
\newblock Boosting algorithms as gradient descent.
\newblock In {\em Proceedings of the 12th International Conference on Neural
  Information Processing Systems}, NIPS'99, pages 512--518, Cambridge, MA, USA,
  1999. MIT Press.

\bibitem{Meisner11}
David Meisner, Christopher~M. Sadler, Luiz~Andr{\'e} Barroso, Wolf-Dietrich
  Weber, and Thomas~F. Wenisch.
\newblock Power management of online data-intensive services.
\newblock In {\em Proceedings of the 38th annual international symposium on
  Computer architecture}, pages 319--330, 2011.

\bibitem{Ousterhout13}
Kay Ousterhout, Patrick Wendell, Matei Zaharia, and Ion Stoica.
\newblock Sparrow: Distributed, low latency scheduling.
\newblock In {\em Proceedings of SOSP}. Farminton, PA, 2013.

\bibitem{qu2018auto}
Chenhao Qu, Rodrigo~N Calheiros, and Rajkumar Buyya.
\newblock Auto-scaling web applications in clouds: A taxonomy and survey.
\newblock {\em ACM Computing Surveys (CSUR)}, 51(4):73, 2018.

\bibitem{Rahman19}
Joy Rahman and Palden Lama.
\newblock Predicting the end-to-end tail latency of containerized microservices
  in the cloud.
\newblock In {\em {IEEE} International Conference on Cloud Engineering, {IC2E}
  2019, Prague, Czech Republic, June 24-27, 2019}, pages 200--210. {IEEE},
  2019.

\bibitem{GoogleTrace}
Charles Reiss, Alexey Tumanov, Gregory Ganger, Randy Katz, and Michael Kozych.
\newblock Heterogeneity and dynamicity of clouds at scale: Google trace
  analysis.
\newblock In {\em Proceedings of SOCC}. 2012.

\bibitem{lime}
Marco~Tulio Ribeiro, Sameer Singh, and Carlos Guestrin.
\newblock "why should {I} trust you?": Explaining the predictions of any
  classifier.
\newblock In {\em Proceedings of the 22nd {ACM} {SIGKDD} International
  Conference on Knowledge Discovery and Data Mining, San Francisco, CA, USA,
  August 13-17, 2016}, pages 1135--1144, 2016.

\bibitem{nr}
Ryan~A. Rossi and Nesreen~K. Ahmed.
\newblock The network data repository with interactive graph analytics and
  visualization.
\newblock In {\em AAAI}, 2015.

\bibitem{rzadca2020autopilot}
Krzysztof Rzadca, Pawel Findeisen, Jacek Swiderski, Przemyslaw Zych, Przemyslaw
  Broniek, Jarek Kusmierek, Pawel Nowak, Beata Strack, Piotr Witusowski, Steven
  Hand, and John Wilkes.
\newblock Autopilot: workload autoscaling at google.
\newblock In {\em Proceedings of the Fifteenth European Conference on Computer
  Systems}, pages 1--16, 2020.

\bibitem{omega13}
Malte Schwarzkopf, Andy Konwinski, Michael Abd-El-Malek, and John Wilkes.
\newblock Omega: flexible, scalable schedulers for large compute clusters.
\newblock In {\em Proceedings of EuroSys}. Prague, Czech Republic, 2013.

\bibitem{Cloudscale}
Zhiming Shen, Sethuraman Subbiah, Xiaohui Gu, and John Wilkes.
\newblock Cloudscale: elastic resource scaling for multi-tenant cloud systems.
\newblock In {\em Proceedings of SOCC}. Cascais, Portugal, 2011.

\bibitem{Sriraman18}
Akshitha Sriraman and Thomas~F. Wenisch.
\newblock {\textmu}tune: Auto-tuned threading for {OLDI} microservices.
\newblock In {\em 13th {USENIX} Symposium on Operating Systems Design and
  Implementation ({OSDI} 18)}, pages 177--194, Carlsbad, CA, October 2018.
  {USENIX} Association.

\bibitem{sriraman2018mu}
Akshitha Sriraman and Thomas~F Wenisch.
\newblock usuite: A benchmark suite for microservices.
\newblock In {\em 2018 IEEE International Symposium on Workload
  Characterization (IISWC)}, pages 1--12. IEEE, 2018.

\bibitem{suresh2017distributed}
Lalith Suresh, Peter Bodik, Ishai Menache, Marco Canini, and Florin Ciucu.
\newblock Distributed resource management across process boundaries.
\newblock In {\em Proceedings of the 2017 Symposium on Cloud Computing}, pages
  611--623. ACM, 2017.

\bibitem{thrift}
Apache thrift.
\newblock \url{https://thrift.apache.org}.

\bibitem{Torque}
Torque resource manager.
\newblock \url{http://www.adaptivecomputing.com/products/open-source/torque/}.

\bibitem{Urgaonkar05}
Bhuvan Urgaonkar, Giovanni Pacifici, Prashant Shenoy, Mike Spreitzer, and Asser
  Tantawi.
\newblock An analytical model for multi-tier internet services and its
  applications.
\newblock {\em SIGMETRICS Perform. Eval. Rev.}, 33(1):291–302, June 2005.

\bibitem{Borg}
Abhishek Verma, Luis Pedrosa, Madhukar~R. Korupolu, David Oppenheimer, Eric
  Tune, and John Wilkes.
\newblock Large-scale cluster management at {Google} with {Borg}.
\newblock In {\em Proceedings of the European Conference on Computer Systems
  (EuroSys)}, Bordeaux, France, 2015.

\bibitem{powerchief}
Hailong Yang, Quan Chen, Moeiz Riaz, Zhongzhi Luan, Lingjia Tang, and Jason
  Mars.
\newblock Powerchief: Intelligent power allocation for multi-stage applications
  to improve responsiveness on power constrained cmp.
\newblock In {\em Proceedings of the 44th Annual International Symposium on
  Computer Architecture}, ISCA ’17, page 133–146, New York, NY, USA, 2017.
  Association for Computing Machinery.

\bibitem{zhou2018overload}
Hao Zhou, Ming Chen, Qian Lin, Yong Wang, Xiaobin She, Sifan Liu, Rui Gu,
  Beng~Chin Ooi, and Junfeng Yang.
\newblock Overload control for scaling wechat microservices.
\newblock In {\em Proceedings of the ACM Symposium on Cloud Computing}, pages
  149--161. ACM, 2018.

\end{thebibliography}

%\newpage
%\appendix
%\input{Appendix.tex}

\end{document}